\documentclass[11pt]{article}
\usepackage[dvips]{graphicx}
\usepackage{graphics}
\usepackage{ifthen}
\usepackage{amsfonts}
\usepackage{amsmath}
\usepackage{epsfig}
\usepackage{youngtab}
\usepackage{multirow}
\usepackage[errorshow]{tracefnt}
 \newfont{\bbbold}{msbm10 scaled \magstep1}

 \def\cL{{\cal L}}

 \def\cO{{\cal O}}

 \newfont{\goth}{eufm10 scaled \magstep1}

 \def\a{\alpha}
 \def\b{\beta}
 \def\c{\gamma}\def\C{\Gamma}
 \def\d{\delta}
 \def\e{\epsilon}\def\vare{\varepsilon}
 
 \def\h{\eta}
 \def\k{\kappa}
 \def\l{\lambda}\def\L{\Lambda}
 \def\m{\mu}

 \def\s{\sigma}\def\S{\Sigma}
 \def\t{\tau}
 \def\th{\theta}

 \def\O{\Omega}
 \def\del{\partial}
 \def\ua{\underline{\alpha}}
 \def\ub{\underline{\phantom{\alpha}}\!\!\!\beta}
 \def\uc{\underline{\phantom{\alpha}}\!\!\!\gamma}
 \def\um{\underline{\mu}}
 \def\ud{\underline\delta}
 
 \def\una{\underline a}\def\unA{\underline A}
 \def\unb{\underline b}\def\unB{\underline B}
 \def\unc{\underline c}\def\unC{\underline C}
 \def\unD{\underline D}

 \def\unm{\underline m}\def\unM{\underline M}
 \def\unN{\underline N}

 \def\unH{\underline{H}}

 \def\xz{\times}

 \def\nab{\nabla}

 \def\del{\partial}

 \def\be{\begin{equation}}\def\ee{\end{equation}}
 \def\bea{\begin{eqnarray}}\def\eea{\end{eqnarray}}
 \def\ba{\begin{array}}\def\ea{\end{array}}
\def\bc{\begin{cases}}\def\ec{\end{cases}}
 \let\la=\label
  \def\nn{\nonumber}
 \def\ft#1#2{{\textstyle{{\scriptstyle #1}\over {\scriptstyle #2}}}}

 \newcommand{\eq}[1]{(\ref{#1})}

\newcommand{\hoch}[1]{$\, ^{#1}$}
\newcommand{\kings}{\it\small Department of Mathematics, King's College,
London, UK}
\newcommand{\upp}{\it\small Department of Theoretical Physics
Uppsala University, Sweden}

\begin{document}

 \thispagestyle{empty}

 \hfill{KCL-MTH-03-13}

 \hfill{ITP-UU-03-11}

  \hfill{\today}

 \vspace{20pt}

 \begin{center}
 {\Large{\bf The deformed M2-brane}}
 \vspace{30pt}

 {P.S. Howe\hoch{1}, S.F. Kerstan\hoch{1}, U. Lindstr\"om\hoch{2}
 and D. Tsimpis\hoch{1}}

 \vspace{60pt}

 \end{center}

\begin{itemize}
\item [$^1$] \kings \item [$^2$] \upp
\end{itemize}

 {\bf Abstract}

The superembedding formalism is used to study correction terms to
the dynamics of the M2 brane in a flat background.  This is done by deforming the standard 
embedding constraint. It is shown
rigorously that the first such correction occurs at dimension
four. Cohomological techniques are used to determine this
correction explicitly. The action is derived to quadratic order in
fermions, and the modified $\k$-symmetry transformations are
given.

 {\vfill\leftline{}\vfill \vskip  10pt

 \baselineskip=15pt \pagebreak \setcounter{page}{1}

%%%%%%%%%%%%%%%%%%%%%%%%%%%%%%%%%%%%%%%%%%%%%%%%%%%%%%%%%%%
%%%%%%%%%%%%%%%%%%%%
%
%   Document
%
%%%%%%%%%%%%%%%%%%%%%%%%%%%%%%%%%%%%%%%%%%%%%%%%%%%%%%%%%%
%%%%%%%%%%%%%%%%%%%%%

\section{Introduction}

An interesting problem in string/M theory is to consider
higher-order  corrections to the effective dynamics of various
branes. A number of results have been obtained in the purely
bosonic sector, particularly for D-branes,
\cite{Andreev:1988cb,Bachas:1999um,Fotopoulos:2001pt,
Wyllard:2000qe,Bilal:2001hb,Wyllard:2001ye,Koerber:2002zb}.\footnote{For
some earlier work on higher derivative corrections see
\cite{Lindstrom:1987ps,Lindstrom:1988fr}.} Some of these 
have been supersymmetrised, for example the $\del^4 F^4$ and other
terms in D-brane actions \cite{Collinucci:2002gd,Drummond:2003ex}, but
it has so far proven difficult to obtain Lagrangians or equations of motion
with fully-fledged kappa-symmetry. Some kappa-symmetric results have been obtained
for higher-derivative terms for superparticles
\cite{Ivanov:1990cb,Gauntlett:1991dw,Ivanov:1991ub} and
superstrings \cite{Curtright:1987mr}, for coincident D0-branes in
\cite{Sorokin:2001av}, and for branes in lower-dimensional
spacetimes in \cite{Howe:2001wc,Drummond:2002kg}. The last two
papers made use of the superembedding formalism \cite{stvz,s}
which offers a systematic way of incorporating kappa-symmetry
whilst maintaining manifest target space supersymmetry. In this
paper we shall examine the leading correction terms for the
M2-brane using this formalism. We shall derive the action up to
terms quadratic in the fermions as well as the modified
kappa-symmetry transformations.

The two examples studied in \cite{Howe:2001wc,Drummond:2002kg},
namely a membrane in four dimensions and a set of coincident
membranes in three dimensions, were relatively easy to analyse
because the worldvolume multiplets in both cases are off-shell.
This means that the standard superembedding constraints do not
need to be altered - one only needs to find higher-order
Lagrangians which can be constructed using standard methods.
However, for the 1/2 BPS branes in string theory and M-theory the
worldvolume multiplets are maximally supersymmetric matter or
gauge multiplets and it is not known how to find auxiliary fields
for these or even if they exist. Indeed, the standard
superembedding constraint leads directly to the equations of
motion of the brane in most cases \cite{Howe:1996mx,Howe:1996yn}.
This means that we need to find new techniques to discuss
higher-derivative interactions for these branes. The purpose of
the present paper is to show how this can be carried out in a
perturbative fashion for the case of the M2-brane. The basic idea
is to deform the superembedding constraint in such a way that the
field content of the brane theory is unchanged. The torsion
identities, which we describe in detail below, then impose
consistency requirements on the deformation which can be analysed
order by order in a parameter $\ell$ which has the dimensions of
length and which could be the Planck length or the string length
$\sqrt{\a'}$. In fact, what we end up with is a cohomology problem
which is similar in spirit to the cohomological problems
encountered in the analysis of higher-order corrections to
maximally supersymmetric field theories including supergravity
\cite{cnta,cntb,cntc,cntd,Howe:2003cy}. However, the cohomology
problem considered here is manifestly covariant with respect to
target space supersymmetry and local worldvolume supersymmetry
($\k$-symmetry); if we were to work in the static gauge the
formalism would automatically have ``$N=2$'' worldvolume
supersymmetry (actually $d=3, N=16$) non-linearly realised.

We shall limit ourselves to the case of  a flat target superspace
in this paper for simplicity, although there will also be
corrections involving the target space curvature.  We show
rigorously that supersymmetry does not allow any deformations with
three or fewer powers of $\ell$, a result which is to be expected
in view of the results that have been obtained for D-branes. We
then analyse in some detail the lowest-order non-trivial
deformation at $\ell^4$. This turns out to be rather complicated
and involves up to seven powers in fields. We show that there is a
unique deformation with at most three fields. We believe that it
is unlikely that there are other independent deformations, because
of the number of constraints that have to be satisfied,  but we
have not completed the proof of this assertion. Knowledge of the
cubic terms in the deformation is sufficient to determine the
modified Lagrangian to quadratic order in fermions, and thus to
determine the full purely bosonic part of the action at order
$\ell^4$. The result is, as expected, a Lagrangian which is
quartic in the extrinsic curvature.

The organisation of the paper is as follows.  In the next section
we briefly review the superembedding formalism and the relation of
worldvolume supersymmetry  to $\k$-symmetry. In section three we
solve the torsion identities for the membrane to first order in
the deformation field $\psi$ and summarise the lowest order
results which give the standard dynamics. In section four we
introduce the relevant spinorial cohomology groups for deformed
superembeddings. In section five we discuss the constraints on the
deformation and solve the relevant cohomology problem up to order
$\ell^4$ in section six. In section seven we construct the action,
up to terms quadratic in the fermions, and in section eight we
discuss the modified $\k$-symmetry transformations.

\section{Superembeddings}

The superembedding formalism was pioneered in the context of
superparticles in three and four dimensions by Sorokin, Tkach,
Volkov and Zheltukhin \cite{stvz}, and has been applied by these
and other authors to various other branes; for a review see
\cite{s}.\footnote{For an earlier attempt to use source and target
superspaces, see, e.g., \cite{Gates:1985vk}.} In
\cite{Howe:1996mx} it was shown that the formalism can be applied
to arbitrary branes including those with various types of
worldvolume gauge fields, and it was then used to construct the
full non-linear equations of motion of the M-theory 5-brane in an
arbitrary supergravity background \cite{Howe:1996yn}.

 We consider a superembedding $f:M\rightarrow \unM$, where $M$ is
 the worldvolume of the brane and $\unM$ is the target space.
 Our index conventions are as follows: coordinate
 indices are taken from the middle of the alphabet with capitals
 for all, Latin for bosonic and Greek for fermionic, $M=(m,\m)$,
 tangent space indices are taken in a similar fashion
 from the beginning of the
 alphabet so that $A=(a,\a)$. We denote the coordinates of $M$
 by $z^M=(x^m,\th^{\m})$.
 The distinguished tangent space
 bases are related to coordinate bases by means of the
 supervielbein, $E_M{}^A$, and its inverse $E_A{}^M$.
 We use exactly the same notation
 for the target space but with all of the indices underlined.
 Indices for the normal bundle are denoted by primes, so that
 $A'=(a',\a')$. We shall occasionally group together tangent and
 normal indices and denote them by a bar, $\bar a=(a,a')$ etc.

 The embedding matrix is the derivative of $f$ referred to the
 preferred tangent frames, thus

  \be
  E_A{}^{\unA}:= E_A{}^M\del_M z^{\unM} E_{\unM}{}^{\unA}
  \la{2.1}
  \ee

In addition, we can specify a basis $\{E_{A'}\}$ for the normal
bundle in terms of the target space basis by means of the normal
matrix $E_{A'}{}^{\unA}$.

 The basic embedding condition is

  \be
  E_{\a}{}^{\una}=0
  \la{2.1.1}
  \ee

Geometrically this states that the odd tangent space of the brane
is a subspace of the odd tangent space of the target superspace at
any point on the brane. To see the content of this constraint we
can consider a linearised embedding in a flat target space in the
static gauge. This gauge is specified by identifying the
coordinates of the brane with a subset of the coordinates of the
worldvolume, so that

 \bea
 x^{\una}&=&\bc x^a  \\ x^{a'}(x,\th)\ec  \\
 \th^{\ua}&=&\bc  \th^{\a}\\ \th^{\a'}(x,\th) \ec
 \la{2.2}
 \eea

Since

 \be
 E^{\una}=d x^{\una} -{i\over2}d\th^{\ua} (\C^{\una})_{\ua\ub}
 \th^{\ub}
 \la{2.3}
 \ee

in flat space, it is easy to see, to first order in the transverse
fields, that the embedding condition implies that

 \be
 D_{\a} X^{a'}= i(\C^{a'})_{\a\b'} \th^{\b'}
 \la{2.4}
 \ee

where

 \be
 X^{a'}=x^{a'} + {i\over2} \th^{\a} (\C^{a'})_{\a\b'} \th^{\b'}
 \la{2.5}
 \ee

{}From this equation the nature of the worldvolume multiplet
specified by the embedding condition can be determined. Depending
on the dimensions involved, this multiplet can be one of three
types: (i) on-shell, i.e. the multiplet contains only physical
fields and these fields satisfy equations of motion, (ii)
Lagrangian off-shell, meaning that the multiplet also contains
auxiliary fields (in most cases), and that the equations of motion
of the physical fields are not satisfied, although they can be
derived from a superfield action, and (iii) underconstrained,
which means that further constraints are required to obtain
multiplets of one of the first two types. For thirty-two target
space supersymmetries, the multiplets are always of type (i) or of
type (iii), whereas for sixteen or fewer supersymmetries all three
types of multiplet can occur, with type (iii) arising typically
for cases of low even codimension.

In this paper we shall be concerned with the membrane in $D=11$
superspace, i.e. the M2-brane. This has been discussed as a
superembedding in \cite{Bandos:1995zw}. The worldvolume is
$d=3,N=8$ superspace, and the worldvolume multiplet is an on-shell
scalar multiplet (type (i)). This can be seen from \eq{2.5}. The
leading component of the superfield $X^{a'}$ describes eight
scalar fields on the worldvolume, while the leading component of
$\th^{\a'}$ describes eight spin one-half fields. Equation
\eq{2.5} implies that there are no further independent component
fields and that the scalars and spinors satisfy the usual
equations of motion for free massless fields.

In order to determine the consequences of the superembedding
condition in the non-linear case, and to find the induced
supergeometry on the brane, one uses the torsion equation which is
the pull-back  of the equation defining the target space torsion
two-form. This reads

 \be
 2\nab_{[A} E_{B]}{}^{\unC} + T_{AB}{}^C E_C{}^{\unC} =
 (-1)^{A(B+\unB)}E_B{}^{\unB}
 E_A{}^{\unA} T_{\unA\unB}{}^{\unC}
 \la{2.6}
 \ee

In this equation the covariant derivative acts on both worldvolume
and target space tensor indices. The latter are taken care of by
the pull-back of the target space connection, while the
worldvolume connection can be chosen in a variety of different
ways. One also has to parametrise the embedding matrix. Having
done all this, one can work through the torsion equation starting
at dimension zero. In this way the consequences of the embedding
condition \eq{2.1.1} can be worked out in a systematic and
covariant fashion.

The equations of the component or Green-Schwarz formalism can be
obtained by taking the leading ($\th=0$) components of the various
equations that describe the brane multiplet. A key feature of the
formalism is that these component equations are guaranteed to be
$\k$-symmetric because $\k$-symmetry can be identified with the
leading term in a  worldvolume local supersymmetry transformation.
We recall briefly how this works \cite{stvz}. Let $v^M$ be a
worldvolume vector field generating an infinitesimal
diffeomorphism. If we write the superembedding in local
coordinates as $f^{\unM}=z^{\unM}(X)$, then the effect of such a
transformation on $\underline{z}(z)$ is

 \be
 \d z^{\unM}= v^M\del_M z^{\unM}
 \la{2.7}
 \ee

If we express this in the preferred bases we find

 \be
 \d z^{\unA}:=\d z^{\unM}E_{\unM}{}^{\unA}=v^A E_A{}^{\unA}
 \la{2.8}
 \ee

Now if we take an odd worldvolume diffeomorphism, $v^a=0$, and use
the embedding condition \eq{2.1.1} we get

 \bea
 \d z^{\una} &=& 0\\
 \d z^{\ua} &=& v^{\a} E_{\a}{}^{\ua}
 \la{2.9}
 \eea

This can be brought to the more usual $\k$-symmetric form if we
define

 \be
 \k^{\ua}:= v^{\a} E_{\a}{}^{\ua}
 \la{2.10}
 \ee

and note that it satisfies

 \be
 \k^{\ua}=\k^{\ub}
 P_{\ub}{}^{\ua}:={1\over2}\k^{\ub}(1+\C)_{\ub}{}^{\ua}
 \la{2.11}
 \ee

where $P$ is the projection operator onto the worldvolume subspace
of the odd tangent space of the target superspace. We can always
write $P=1/2(1+\C)$, and so $\C$ is computable in terms of the
embedding matrix. Substituting this into \eq{2.9} we recover the
normal form of $\k$-symmetry transformations. (Strictly, we should
evaluate this equation at $\th=0$ to get the correct component
form.) The explicit form of $P$ is

 \be
 P_{\ua}{}^{\ub}=(E^{-1})_{\ua}{}^{\c} E_{\c}{}^{\ub}
 \la{2.12}
 \ee

where the inverse is taken in the fermionic tangent space (of
$\unM$).

For the case of the supermembrane in $D=11$, this procedure yields
the equations of motion found in \cite{Bergshoeff:1987cm}. To find
higher derivative corrections to these equations it follows that
we shall need to amend the basic embedding condition \eq{2.1.1}.
We note that a simple consequence of this is that the
$\k$-symmetry transformations will no longer have the standard
characteristic form for which $\d z^{\una}=0$.

\section{The torsion identities}

In this section we shall study the torsion identities in the case
that the embedding condition is relaxed. We shall take the target
space to be flat so that the only non-vanishing component of the
target space torsion is

\be T_{\ua\ub}{}^{\una}=-i\left(\C^{\una}\right)_{\ua\ub},
\la{3.1} \ee

We can parametrise the embedding matrix as follows:

\bea E_{\a}{}^{\una}=& \psi_\a{}^{a'} u_{a'}{}^{\una}\qquad
&E_{\a}{}^{\ua}=u_{\a}{}^{\ua}
 + h_{\a}{}^{\a'} u_{a'}{}^{\ua}\la{3.2}\\
E_a{}^{\una}=&u_a{}^{\una},\qquad &E_a{}^{\ua} =\Lambda_a{}^{\a '}
u_{\a'}{}^{\ua} \la{3.3} \eea

while the normal matrix can be chosen to have the form

\bea
E_{\a'}{}^{\una}=& 0,\qquad &E_{\a'}{}^{\ua}=u_{\a'}{}^{\ua}\la{3.4}\\
E_{a'}{}^{\una}=&u_{a'}{}^{\una}, \qquad &E_{a'}{}^{\ua}=0
\la{3.5} \eea

Here, the $32\xz 32$ matrix $u_{\overline\a}{}^{\
\ua}:=(u_{\a}{}^{\ua},\,u_{\a'}{}^{\ua})$
 is an element of $Spin(1,10)$
while the matrix $u_{\overline a}{}^{\
\una}:=(u_a{}^{\una},\,u_{a'}{}^{\una})$ is the corresponding
element of $SO(1,10)$.  The dimensions, in units of mass, of the
various components of the embedding matrix are given by

\be [E_{\a}{}^{\ua}]=[E_{a}{}^{\una}]=0,\qquad
[E_{\a}{}^{\una}]=-[E_{a}{}^{\ua}]=-\ft{1}{2} \ee

and similarly for the normal matrix. We can always bring these
matrices into the above forms by a suitable choice of the bases of
the even and odd tangent spaces on the worldvolume. We can now
plug this form of the embedding matrix into the torsion identity

\be \nabla_A E_B{}^{\unC}-(-)^{AB}\nabla_B
E_A{}^{\unC}+T_{AB}{}^{C} E_C{}^{\unC}=(-)^{A(B+\unB
)}E_B{}^{\unB} E_A{}^{\unA} T_{\unA\unB}{}^{~~\unC} \ee

and work out the consequences. As we noted earlier, the covariant
derivative here acts on both target space and worldvolume indices,
but since the target space is flat we only need a worldvolume
connection. This can be specified either by imposing some
constraints on the worldvolume torsion tensor or by choosing a
connection which is natural in the superembedding context. We
shall use the latter. Given the matrix $u$ we can define the set
of one-forms

\be X_{A}:= (\nab_A u) u^{-1} \la{} \ee

If we set

\be X_{A,b}{}^c = X_{A,b'}{}^{c'}=0 \la{} \ee

then we fix connections for the tangent and normal bundles in a
standard fashion. Note that

\be X_{A,\bar\b}{}^{\bar\c}={1\over4} (\C^{\bar b\bar
c})_{\bar\b}{}^{\bar\c} X_{A,\bar b\bar c} \la{} \ee

The torsion identities yield the following results for a flat
target superspace:

dim 0

 \bea
 T_{\a\b}{}^c&=& -i(\C^c)_{\a\b}+2\psi_{(\a}{}^{d'}X_{\b),d'}{}^c\\
 \nab_{(\a}\psi_{\b)}{}^{c'}&=&-ih_{(\a}{}^{\c'}(\C^{c'})_{\b)\c'}
 +\left(-T_{\a\b}{}^{\c}\psi_{\c}{}^{c'}\right)\la{27}
 \eea

dim 1/2

 \bea
 T_{\a b}{}^c&=&\psi_{\a}{}^{d'}
 X_{b,d'}{}^c+i\L_b{}^{\b'}h_{\a}{}^{\c'}(\C^c)_{\b'\c'}\\
 X_{\a
 b}{}^{c'}&=&\nab_b\psi_{\a}{}^{c'}+i\L_b{}^{\b'}(\C^{c'})_{\a\b'}\\
 T_{\a\b}{}^{\c}&=&-2h_{(\a}{}^{\d'}X_{\b),\d'}{}^{\c}\\
 2\nab_{(\a}h_{\b)}{}^{\c'}&=&-2X_{(\a,\b)}{}^{\c'}-T_{\a\b}{}^c\L_c{}^{\c'}
 -\left(T_{\a\b}{}^\d h_\d{}^{\c'}\right)
 \eea

dim 1

 \bea
 T_{ab}{}^c&=&i\L_a{}^{\a'}\L_b{}^{\b'}(\C^c)_{\a'\b'}\\
 2X_{[ab]}{}^{c'}&=&-T_{ab}{}^{\c}\psi_{\c}{}^{\c'}\\
 T_{a\b}{}^\c&=&-\L_a{}^{\d'}X_{\b,\d'}{}^\c
 -h_\b{}^{\d'}X_{a,\d'}{}^\c\\
 X_{a,\b}{}^{\c'}&=&\nab_\b \L_a{}^{\c'}-\nab_a h_\b{}^{\c'}
 \eea

dim 3/2

 \bea
 T_{ab}{}^{\c}&=& 2\L_{[a}{}^{\d'}X_{b],\d'}{}^\c  \\
 2\nab_{[a}\L_{b]}{}^{\c'}&=&-T_{ab}{}^c \L_c{}^{\c'}-T_{ab}{}^\d
 h_\d{}^{\c'}
 \eea

The terms in brackets will be irrelevant in what follows because
they will turn out to be zero at first order in the deformation
parameter as we shall shortly see.

\subsection{Zeroth order}

In the zeroth order theory the standard embedding condition
$E_\a{}^{\una}=0$ holds, so that $\psi_{\a}{}^{c'}=0$. If we
substitute this into the above equations we find that
$h_{\a}{}^{\b'}=0$ while the induced torsion is

 \be
 T_{\a\b}{}^c=-i(\C^c)_{\a\b}
 \ee

at dimension zero and vanishes at dimension one-half. The
dimension one and three-halves components are given by the
corresponding  equations above.

For $X$ we find, at dimension one-half,

 \bea
 X_{\a,b}{}^{c'}&=&i\L_b{}^{\b'}(\C^{c'})_{\a\b'}\\
 2X_{(\a\b)}{}^{\c'}&=&i(\C^c)_{\a\b}\L_c{}^{\c'}
 \eea

It is easy to show that these equations imply that

 \be
 (\C^a)_{\a'\b'}\L_a{}^{\b'}=0
 \la{deq}
 \ee

In the linearised theory $\L_a{}^{\b'}\sim \del_a \th^{\b'}$, so
that we can identify \eq{deq} as the equation of motion of the
fermion field in the worldvolume multiplet. At dimension one we
have

 \bea
 X_{[ab]}{}^{c'}&=&0\\
 X_{a,\b}{}^{\c'}&=&\nab_\b \L_a{}^{\c'}
 \eea

Using the fermion equation of motion we find the scalar equation
of motion

 \be
 \h^{ab}X_{a,b}{}^{c'}=0
 \ee

Indeed, in the linearised theory, $X_{ab}{}^{c'}\sim \del_a\del_b
X^{c'}$, so we get the massless Klein-Gordon equation. Finally, a
short calculation gives the supersymmetry variation of
$X_{ab}{}^{c'}$:

 \be
 \nabla_{\a i}X_{b,c}{}^{d'}=-i(\s^{d'}\nabla_{(b}\L_{c)})_{\a i}
+{1\over 2}(\L_b\c^e\otimes
\tilde{\s}^{f'd'}\L_c)(\s_{f'}\L_e)_{\a i}
 \ee

\subsection{First order}

A first order deformation of the theory will involve the presence
of a non-vanishing $\psi$ field which we can take to be of the
form of a dimensionful parameter $\b$, say, multiplied by a
function of the physical fields $\L_a{}^{\b'}$ and $X_{ab}{}^{c'}$
as well as derivatives. We shall write this schematically as

 \be
 \psi=\b f(\L,X,\del)
 \ee

Furthermore, $\psi$ is subject to the constraint

 \be
 \nab_{(\a}\psi_{\b)}{}^{c'}=-ih_{(\a}{}^{\c'}(\C^{c'})_{\b)\c'}
 \la{47}
 \ee

Note that, since $\psi,h$ and $T_{\a\b}{}^\c$ are all of order
$\b$ we are allowed to drop the $T_{\a\b}{}^\c \psi_{\c}{}^{c'}$
term from \eq{27}. The problem is then to analyse equation
\eq{47}. This can be done systematically in powers of the length
parameter $\ell$. Note that, given an explicit expression for
$\psi$ in terms of the physical fields, equation \eq{47} allows us
to solve for $h$.

\section{Spinorial cohomology for branes}

The notion of spinorial cohomology \cite{cnta,cntc} is useful for studying both the
space of physical fields in certain supersymmetric theories and
also for studying deformations of the equations of motion. In the
simplest case one studies spinorial $p$-forms, i.e. totally
symmetric $p$-spinors which are $\c$-traceless, together with a
differential operator which is obtained by acting with $D_\a$
followed by symmetrisation and removal of the $\c$-trace. This
defines a cohomology which is isomorphic to pure spinor cohomology
\cite{Berkovits:2000fe,Berkovits:2001rb}. The relevance of the latter to theories
in ten and eleven dimensions is explained by the fact that the
equations of motion can be interpreted in terms of pure spinor
integrability \cite{Howe:mf,Howe:1991bx}. The formalism has been
applied to $D=10$ Yang-Mills theory and supergravities in ten and
eleven dimensions \cite{cnta,cntb,cntc,cntd,Howe:2003cy}.
Moreover, one can also consider vector-valued spinorial cohomology
\cite{cntc}. For branes, the appropriate cohomology is a variant
on the latter.

Since we are interested in first-order deformations, it is
sufficient to work on a worldvolume whose induced supergeometry is
given by the zeroth order theory. We then consider the spaces
$\O^p_{B'}$ and $\O^p_{F'}$. The objects in these spaces are
spinorial $p$-forms on $M$ which take their values either in the
even normal bundle $B'$ or the odd normal bundle $F'$. There is a
natural map $\C:\O^{p-1}_{F'}\rightarrow \O^p_{B'}$ given by

 \be
 h_{\a_1\ldots \a_{p-1}}{}^{\c'}\mapsto -i
 h_{(\a_1\ldots \a_{p-1}}{}^{\c'}(\C^{c'})_{\a_b)\c'}
 \la{equiv}
 \ee

We can therefore form the quotient space
$\hat\O^p_{B'}:=\O^p_{B'}/\C(\O^{p-1}_{F'})$ and construct a
derivative $d_s:\hat\O^p_{B'}\rightarrow\hat\O^{p+1}_{B'}$ which
squares to zero. This derivative is defined by acting with the
spinorial covariant derivative $\nab_\a$ and symmetrising modulo
equivalences of the form of equation \eq{equiv}. In other words,
if $h\in\O^p_{B'}$ represents an equivalence class
$[h]\in\hat\O^p_{B'}$, then $d_s[h]=[\nab h]\in\hat\O^{p+1}_{B'}$
where

 \be
 (\nab h)_{\a_1\ldots \a_{p+1}}{}^{c'}:=\nab_{(\a_1}h_{\a_2\ldots
 \a_{p+1})}{}^{c'}
 \ee

It is easy to see that this definition is independent of the
choice of representative $h$. To see that $d_s^2=0$ we observe
that

 \be
 \nab_{(\a_1}\nab_{\a_2}h_{\a_3\ldots\a_{p+2})}{}^{c'}=
 -{1\over2} R_{(\a_1\a_2,\a_3}{}^{\c} h_{|\c|\a_4\ldots
 \a_{p+2})}{}^{c'}+{1\over2}
 R_{(\a_1\a_2,d'}{}^{c'}h_{\a_3\ldots\a_{p+2})}{}^{d'}
 \la{50}
 \ee

Now, from the first Bianchi identity, we have

 \bea
 R_{(\a_1\a_2\a_3)}{}^\c&=&\nab_{(\a_1}T_{\a_2\a_3)}{}^\c +
 T_{(\a_1\a_2}{}^B T_{B \a_3)}{}^{\c}\nn\\
 &=&-i(\C^b)_{(\a_1\a_2}T_{b\a_3)}{}^{\c}
 \eea

so that the first term on the RHS of \eq{50} is of the form

 \be
 -i(\C^b)_{(\a_1\a_2} k_{b\a_3\ldots \a_{p+2})}{}^{c'}
 \ee

for some $k$. However, this can be written as

 \be
 -ih_{(\a_1\ldots \a_{p+1}}{}^{\c'} (\C^{c'})_{\a_{p+2})\c'}
 \ee

where

 \be
 h_{\a_1\ldots
 \a_{p+1}}{}^{\c'}=(\C^{bc'})_{(\a_1}{}^{\c'}k_{b\a_2\ldots\a_{p+2})c'}
 \ee

and so this term maps to zero in the quotient space. The second
term on the RHS of \eq{50} can be evaluated with the aid of the
Gauss-Codazzi equation which can, in turn, be derived from the
definition of $X_A$. One finds that

 \be
 R_{\a\b c'}{}^{d'}=2X_{(\a c'}{}^e X_{\b) e}{}^{d'}
 \ee

Since $X_{\a b}{}^{c'}=i\L_b{}^{\b'}(\C^{c'})_{\b'\a}$ we have

 \be
 R_{\a\b c'}{}^{d'}=2\L^{e\c'}\L_e{}^{\d'}
 (\C^{c'})_{\c'(\a}(\C^{d'})_{\b)\d'}
 \ee

from which it is easy to see that the second term also gives zero
in the quotient space. This shows that $d_s^2=0$ so that we can
define the cohomology groups $H^p_{B'}:=\ker
d_s\cap\hat\O^p_{B'}/{\rm im} d_s\hat\O^{p-1}_{B'}$. We claim that
a first-order deformation of the dynamics of the membrane is given
by an element of $H^1_{B'}(phys)$, where the notation indicates
that the coefficients should be given by fields constructed from
the physical fields.

To see that this is the appropriate group we need to consider
redefinitions. We shall find the effect of a field redefinition of
the embedding coordinates
\begin{equation}
z^{\unM}\rightarrow z^{\unM}+(\d z)^{\unM} \label{trnsf}
\end{equation}
on the field $\psi_{\a}{}^{c'}$. The transformation (\ref{trnsf})
of the embedding coordinates with the background geometry fixed
may be viewed equivalently as a diffeomorphism of the background
geometry with the embedding coordinates fixed. Taking the latter
point of view, we find that the target-space vielbein transforms
as
$$
\d E_{\unM}{}^{\unA}=(\d z)^{\unN} T_{\unN \unM}{}^{\unA}
+\nab_{\unM}(\d z)^{\unA},
$$
up to a Lorentz transformation on the flat index. The
transformation of the embedding matrix reads
\begin{align}
\d E_A{}^{\unA}&=E_A{}^M\partial_M Z^{\unM} \d E_{\unM}{}^{\unA}\nn\\
&=\nab_{A}(\d z)^{\unA}+E_A{}^{\unC}(\d z)^{\unB}T_{\unB
\unC}{}^{\unA}\nn
\end{align}
Setting $A=\a,~ \unA=\una$ in the above and expressing
$E_{\a}{}^{\una}$ in terms of  $\psi_\a{}^{c'}$, we obtain

 \be
 \d \psi_\a{}^{c'}=\nab_{\a}(\d z)^{c'}+(\d z)^b X_{\a b}{}^{c'} -i
 (\d z)^{\c'}(\C^{c'})_{\c' \a}
 \ee

where $(\d z)^{\overline{c}}:=(\d
z)^{\unc}u_{\unc}{}^{\,\overline{c}}$, etc. Since $\d z$ is of
order $\b$ we can use the zeroth order expression for $X$ and so
we obtain

 \be
 \d \psi_\a{}^{c'}=\nab_{\a}(\d z)^{c'} -i
 h^{\c'}(\C^{c'})_{\c' \a}
 \la{redef}
 \ee

where

 \be
 h^{\c'}:=(\d z)^{\c'}- (\d z)^b \L_b{}^{\c'}
 \ee

Together with the fact that $\psi$ satisfies the constraint
\eq{47}, this proves the contention that the first-order
deformation is indeed given by an element of $H^1_{B'}(phys)$.

We observe that the zeroth order group $H^0_{B'}$  can be
interpreted as the space of physical fields, since a deformation
of the form of \eq{redef} such that the left-hand side vanishes
will leave the basic embedding constraint unchanged. Indeed, the
linearised embedding equation \eq{2.1.1} can also be viewed as
defining an element of this group in flat space.

The discussion given here is pertinent to the M2-brane but can be
generalised to other branes. This may require some technical
modifications since the field $h_\a{}^{\c'}$ does not vanish at
zeroth order in the presence of worldvolume gauge fields; in
particular, this is the case for D-branes and the M5-brane.

\section{Constraints on $\psi$}

We now derive the constraints on $\psi_\a{}^{a'}$ implied by the
dimension--0 torsion identity at linear order in $\b$,
\begin{equation}
\nabla_{(\alpha}\psi_{\beta
)}^{~~c'}=-ih_{(\a}{}^{\c'}(\C^{c'})_{\b)\c'} \label{constraint}
\end{equation}
Note that in the above we have taken into account that
$T_{\alpha\beta}^{~~\c}$ is of order $\b$ as implied by the
dimension--1/2 torsion identity.

The field $\psi_{\a i}{}^{a'}$ (switching to two-step notation)
transforms under the $(1)\times((0001)\otimes(1000))$
representation of $Spin(1,2)\times Spin(8)$. (See the appendix for
representation-theoretic conventions.) Decomposing $\psi$ into
irreducible representations, we find

$$
\psi_{\a i}{}^{a'}\sim (1)\times (1001)\oplus (1)\times (0010)
$$
Explicitly:
\begin{alignat}{3}
\psi_{\a i}{}^{a'}&=\S_{\a i}{}^{a'}&\qquad (1)\times (1001)\nn\\
&+(\s^{a'})_{ij'}\S_\a^{j'} &(1)\times (0010) \label{psxp}
\end{alignat}
{}From \eq{redef} it follows that the trace part in the
decomposition of $\psi_{\a i}^{a'}$ above can be eliminated using
field redefinitions of the embedding coordinates.  We shall
therefore set $\S_\a^{i'}=0$.

Similarly, the field $h_{\a}{}^{\b'} \to h_{\a i}{}^{\b j'}$ (in
two-step notation) transforms under the
$((1)\otimes(1))\times((0001)\otimes(0010))$ representation of
$Spin(1,2)\times Spin(8)$. Decomposing in irreducible
representations, we have

$$
h_{\a i}{}^{\b j'} \sim (0)\times (1000)\oplus (0)\times
(0011)\oplus (2)\times (1000)\oplus (2)\times (0011)
$$
Explicitly
\begin{alignat}{3}
h_{\a i}{}^{\b j'}&=\d_\a{}^\b(\s^{a'})_i{}^{j'}h_{a'} &\qquad (0)\times (1000)\nn\\
&+{1\over 6}\d_\a{}^\b(\s^{a'b'c'})_i{}^{j'}h_{a'b'c'} &(0)\times (0011)\nn\\
&+(\c^a)_\a{}^\b(\s^{a'})_i{}^{j'}h_{aa'} &(2)\times (1000)\nn\\
&+{1\over 6}(\c^a)_\a{}^\b(\s^{a'b'c'})_i{}^{j'}h_{aa'b'c'}
&(2)\times (0011) \label{hexp}
\end{alignat}

In order to analyse (\ref{constraint}) we need to decompose
$\nabla_{\a i} \S_{\b j}{}^{a'}$ into irreducible representations
of $Spin(1,2)\times Spin(8)$,  (remember we have set $\S_\a^{i'}$
to zero). This field transforms under the
$((1)\otimes(1))\times((0001)\otimes(1001))$ representation.
Decomposing into irreducible representations, we have
\begin{alignat}{2}
\nabla_{\a i} \S_{\b j}{}^{a'} &\sim (0)\times (1000)\oplus
(0)\times (0011)\oplus
(0)\times (1100)\nn\\
&\oplus (2)\times (1000)\oplus (2)\times (0011)\oplus (2)\times
(1002)\nn
\end{alignat}
Explicitly,
\begin{alignat}{3}
\nabla_{\a i} \S_{\b j}^{a'} &=\vare_{\a\b} \d_{ij} Y^{a'}
+{1\over 7} \vare_{\a\b} (\s^{a'b'})_{ij}
Y_{b'}&\qquad (0)\times (1000)\nn\\
&+{1\over 2}\vare_{\a\b}(\s_{b'c'})_{ij}Y^{a'b'c'} +{1\over
10}\vare_{\a\b}(\s^{a'b'c'd'})_{ij}Y_{b'c'd'}
&(0)\times (0011)\nn\\
&+{1\over 2}\vare_{\a\b}(\s_{b'c'})_{ij}Y^{b'c';a'} & (0)\times (1100)\nn\\
&+{1\over 120}\vare_{\a\b}(\s_{b'c'd'e'})_{ij}Y^{b'c'd'e';a'}
 &(0)\times (1002)\nn\\
&+(\c^a)_{\a\b} \d_{ij} Y_a{}^{a'} +{1\over 7} (\c^a)_{\a\b}
(\s^{a'b'})_{ij}
Y_{ab'}&\qquad (2)\times (1000)\nn\\
&+{1\over 2}(\c^a)_{\a\b}(\s_{b'c'})_{ij}Y_a{}^{a'b'c'} +{1\over
10}(\c^a)_{\a\b}(\s^{a'b'c'd'})_{ij}Y_{ab'c'd'}
&(2)\times (0011)\nn\\
&+{1\over 2}(\c^a)_{\a\b}(\s_{b'c'})_{ij}Y_a{}^{b'c';a'} & (2)\times (1100)\nn\\
&+{1\over 120}(\c^a)_{\a\b}(\s_{b'c'd'e'})_{ij}Y_a{}^{b'c'd'e';a'}
&(2)\times (1002) \label{yexp}
\end{alignat}
The semi-colon notation here denotes ``hook'' representations, e.g
$a'b';c'$ denotes a traceless tensor which is antisymmetric on
$a'b'$, but not on all three indices. Using
(\ref{hexp},\ref{yexp}) in (\ref{constraint}) we can solve for
$h_{\a i}{}^{\b j'}$ in terms of the $Y$s:
\begin{alignat}{2}
h_{a'}&= -{i\over 7}Y_{a'} \nn\\
h_{a'b'c'}&= -iY_{a'b'c'} \nn\\
h_{aa'}&= iY_{aa'} \nn\\
h_{aa'b'c'}&= -{3i\over 5}Y_{aa'b'c'}. \nn
\end{alignat}
In addition we find two constraints
\begin{alignat}{2}
Y_{a'b';c'}&=0\nn\\
Y_{aa'b'c'd';e'}&=0. \label{cnstr}
\end{alignat}
Finally, the fields $Y_{aa'b';c'}, ~ Y_{a'b'c'd';e'}$ drop out of
(\ref{constraint}) and are therefore left undetermined.

\section{Solving the spinorial cohomology problem}

Let us recapitulate: the deformations of the supersymmetric M2
theory are parametrised by the object $\S^{~~~a'}_{\a i}$ in the
$(1)\times(1001)$ of $Spin(1,2)\times Spin(8)$.  $\S_{\a
i}{}^{a'}$ is not arbitrary, but has to satisfy the two
constraints (\ref{cnstr}). In the language of section 4, such a
$\S^{~~~a'}_{\a i}$ determines an element of the spinorial
cohomology group $H^1_{B'}$. These constraints tell us that the
projection  of the spinor derivative of $\S^{~~~a'}_{\a i}$ onto
the $(0)\times(1100) \oplus (2)\times (1002)$ part should vanish.
This gives us the explicit form of the spinorial derivative $d_s$
for this case:
\begin{equation}
d_s(\S_{\a i}{}^{a'}) \to (\s^{a_1'a_2'})^{ij}\e^{\a\b} \nabla_{\a
i}\S^{b'}_{\b j}\vert_{(1100)} \oplus (\s^{a_1'\dots
a_5'})^{ij}\c^{\a\b}_a \nabla_{\a i}\S^{b'}_{\b j}\vert_{(1002)},
\label{dione}
\end{equation}
where the bars denote the projections onto the indicated
irreducible $Spin(8)$ representations. In addition one has to take
into account the field redefinitions. As explained in the previous
section, these are given by the projection of the spinor
derivative of $(\d z)^{a'}$ onto the $(1)\times(1001)$ part.
Explicitly,
\begin{equation}
d_s(\d z)^{a'} \to \nabla_{\a i}(\d z)^{a'}\vert_{(1001)},
\label{ditwo}
\end{equation}
Our strategy is to view $\S^{~~~a'}_{\a i} ,~ (\d Z)^{a'}$ as
composite operators, given in terms of the world-volume fields
$X_{ab}{}^{c'}, ~\L_a{}^{\a i'}$. At any given order of $\ell$,
one writes down the most general expressions for $\S^{~~~a'}_{\a
i} ,~ (\d z)^{a'}$ allowed by  dimensional analysis and
representation theory. In  determining the most general
expressions  at a given order in $\ell$ one can assume that the
worldvolume fields obey the lowest-order equations. This is
because  $\S^{a'}_{\a i} ,~(\d z)^{a'}$ are already of order $\b$.
In terms of irreducible representations of $Spin(1,2)\times
Spin(8)$,
\begin{align}
\L_{a}{}^{\a i'}&\sim (3)\times (0010)\nn\\
X_{ab}^{c'}&\sim (4)\times(1000)\nn\\
\nabla_{a}\L_{b}^{~\a i'}&\sim (5)\times (0010), ~~~{\rm up ~to
~terms ~of ~the ~form}
~\L^3\nn\\
\nabla_a X_{bc}{}^{d'}&\sim (6)\times(1000),~~~{\rm up ~to ~terms
~of ~the ~form} ~X\L^2\nn
\end{align}

\subsection{Spinorial cohomology at $\b=\ell^2$}

By dimensional analysis $\S^{~~~a'}_{\a i}$ is schematically of
the form\footnote{In this equation and other similar schematic
equations, the symbol $\nab$ indicates the even covariant
derivative $\nab_a$.}
$$
\S\sim \nabla \L\oplus X\L\oplus \L^3.
$$
In terms of irreducible representations of $Spin(1,2)\times
Spin(8)$,
\begin{align}
\L^3&\sim\yng(1,1,1)_{(3)}\times
\yng(3)_{(0010)}\oplus\yng(2,1)_{(3)}\times\yng(2,1)_{(0010)}
\oplus\yng(3)_{(3)}\times\yng(1,1,1)_{(0010)}\nn\\
\nabla \L&\sim (5)\times(0010)\nn\\
X\L&\sim (4)\times(1000)\otimes(3)\times(0010),\nn
\end{align}
where we have taken into account the fact that $\L$ is
anticommuting. One can verify that $(1)\times(1001)$ is {\it not}
contained in the decomposition of the right-hand sides above and
therefore there is no possible composite field $\S^{~~~a'}_{\a i}$
at this order in $\b$. Consequently, the spinorial cohomology is
trivial and there are no possible supersymmetric deformations of
the theory at this order.

\subsection{Spinorial cohomology at $\b=\ell^3$ }

In this section we shall show that $H^1_{B'}(phys)=0$ at order
$\b=\ell^3$. By dimensional analysis $\S^{~~~a'}_{\a i}$ has the
form
\begin{equation}
\S\sim \fbox{$X^2\L$}\oplus\nabla^2 \L\oplus \nabla X\L\oplus
X\nabla\L \oplus X\L^3\oplus  \fbox{$\L^2\nabla\L$}
\oplus\fbox{$\L^5$}. \label{cohthree}
\end{equation}
Analysing this in terms of representations of $Spin(1,2)\times
Spin(8)$ one finds that only the boxed terms contain $(1) \times
(1001)$. The three contributing terms are
\begin{align*}
\L^5 &\sim \tiny \yng(1,1,1,1,1)_{(3)} \times \yng(5)_{(0010)}
\oplus \yng(5)_{(3)} \times \yng(1,1,1,1,1)_{(0010)} \oplus
\yng(2,1,1,1)_{(3)}\times \yng(4,1)_{(0010)}\\
&\tiny \oplus \fbox{$\yng(4,1)_{(3)}\times
\yng(2,1,1,1)_{(0010)}$}\oplus
\fbox{$\yng(3,1,1)_{(3)}\times\yng(3,1,1)_{(0010)}$} \oplus
\yng(2,2,1)_{(3)}\times\yng(3,2)_{(0010)}\\
&\tiny \oplus\yng(3,2)_{(3)}\times \yng(2,2,1)_{(0010)}\oplus\\
X^2\L&\sim \tiny \fbox{$\yng(2)_{(4)}\times\yng(2)_{(1000)}
\otimes (3)\times(0010)$}\oplus
\fbox{$\yng(1,1)_{(4)} \times\yng(1,1)_{(1000)} \otimes (3)\times(0010)$}\\
\L^2\nabla\L &\sim \tiny
\fbox{$\yng(2)_{(3)}\times\yng(1,1)_{(0010)} \otimes
(5)\times(0010)$} \oplus\yng(1,1)_{(3)}
\times\yng(2)_{(0010)} \otimes \tiny{(5)\times(0010)}\\
\end{align*}
Analysing the contents of these products in terms of irreducible
representations, we find that only the boxed expressions contain
the representation of $\S$, i.e. $(1) \times (1001)$. So at order
$\ell^3$ a nontrivial $\S$ must be built from these five
terms and must satisfy the constraints \eq{cnstr}.\\

We now identify the field redefinitions which can be used to
remove some of these terms. Dimensional analysis restricts the
field redefinitions $(\d z)^{a'}$ at order $\ell^3$ to be of the
types $X^2$, $X\L\L$ and $\L^4$. Decomposing these into
irreducible representations one finds that that only $X\L\L$
contains a $Spin(8)$ vector which is a $Spin(1,2)$ scalar as
required, and so the field redefinitions are given as:

$$\L^2X \sim \left(\yng(1,1)_{(3)} \times\yng(2)_{(0010)}\right)
\otimes (4)\times(1000)$$.

This means that we can remove one of the five terms in $\S$ by a
field redefinition.

Next, we need to find the terms which contribute to the
constraints \eq{cnstr}. This can be done either by explicitly
calculating the action of $d_s$ on all five terms, or by
dimensional analysis and representation theory. The latter method
reveals that, schematically

$$Y \sim X \L \nab \L, \L^2 \nab X \hspace{1mm} \text{ and }
\hspace{1mm} \L^4 X~.$$

Figure \ref{order3} illustrates how the field redefinitions
generate contributions to $\S$ and how $\S$ generates
contributions to the constraints.
\leavevmode
\begin{figure}[h] \label {order3}
\begin{center}
\input{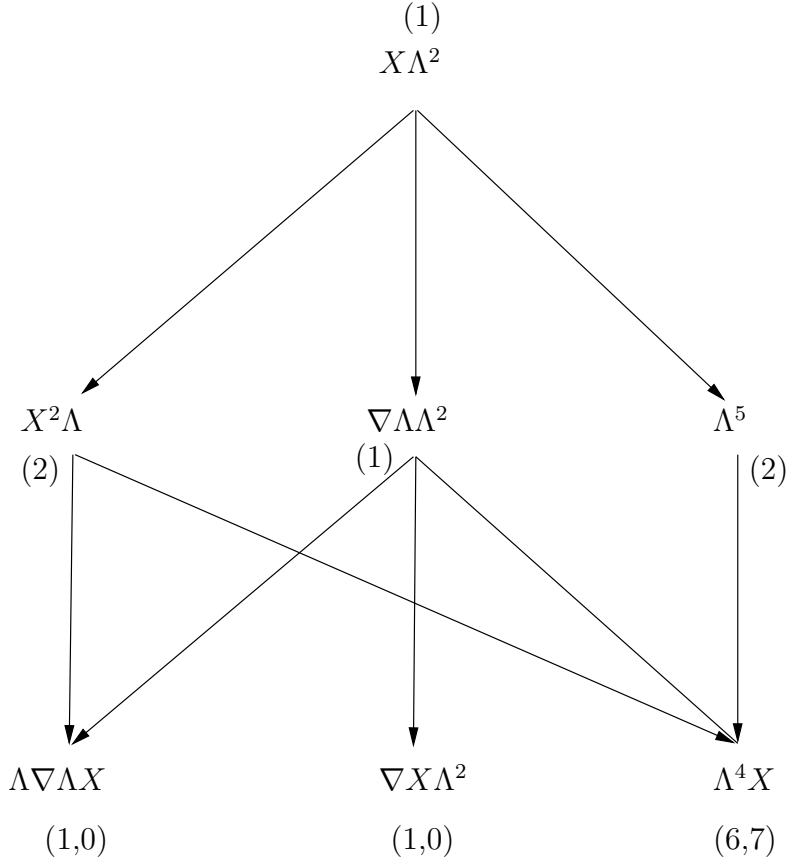}
\end{center}
\caption{Spinorial cohomology at  $\ell^3$. The three rows
contain, schematically, all possible terms in $(\d z)^{a'}$ (field
redefinitions),  $\S^{~~~a'}_{\a i}$ (traceless part of $\psi$),
$Y_{a'b';c'}$ and $Y_{aa'b'c'd';e'}$ (constraints). Multiplicities
are denoted by the numbers in parentheses. In the third row  the
first number in each parenthesis denotes the multiplicity of
$Y_{a'b';c'}$ and the second the multiplicity of
$Y_{aa'b'c'd';e'}$. The arrows from the first to second, and
second to third, rows indicate the action of $d_s$.}
\end{figure}
We can choose to remove the $\L^3 \nab\L$-type term, and are then
left with four terms that must not violate the constraints, i.e.
contribute to the $(0) \times (1100)$ or $(2) \times (1002)$. To
show that any linear combination of these terms, that is any
content of $\psi$ at this order in $\ell$, would violate the
constraints, we need to go into more detail. We break up the
possible terms for $\S$ and those for the constraints $Y$ into
products of irreducible representations (which one can derive by
dimensional analysis and representation theory as before). This is
illustrated in figure \ref{order3.2}.
\begin{figure} \label{order3.2}
\begin{center}
\input{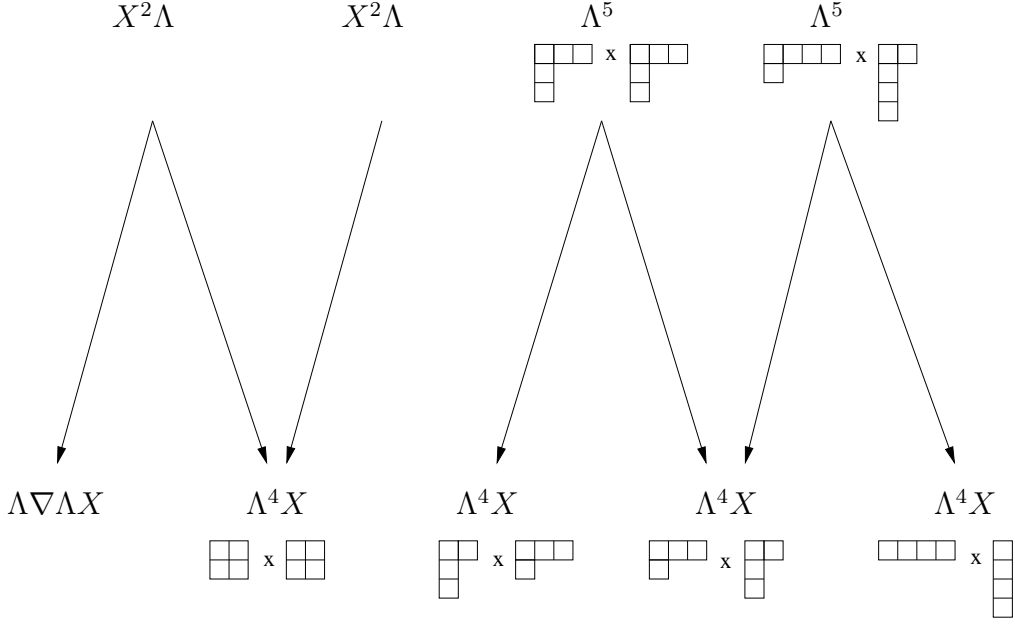}
\end{center}
\caption{The first row represents possible terms for $\S$ and the
second row those for $Y$. The arrows indicate the action of $d_s$
(the spinorial derivative). Some arrows are suppressed. The
products of Young-Tableaux represent plethysms of $\L$, so that
the left factor is a tableau of the (3) of Spin(1,2) and the right
factor is a tableau of the (0010) of Spin(8).}
\end{figure}
As indicated in the figure, after using the field redefinition, we
are left with two different $X^2\L$ and two different $\L^5$ terms
in $\S$. We can choose a basis for the $X^2\L$ such that only one
of them contributes to the forbidden $\L \nab\L X$ (arrows
indicating its contributions to other representation in $Y$ are
suppressed). The latter cannot be generated from either of the
$\L^5$ terms. Therefore, the first $X^2\L$-term would violate the
constraints and cannot be present. An explicit calculation then
shows that the remaining $X^2\L$ term  contributes non-trivially
to the $X\L^4$ with a (2,2) plethysm of the $\L$. Again, neither
of the $\L^5$ terms can cancel this, and so the second $X^2\L$
term must also be trivial (again, arrows indicating its
contributions to other representations in $Y$ are suppressed).
Finally, as the arrows in the figure indicate, the $\L^5$ terms
cannot remove each other's contributions towards $Y$. This shows
that there is no linear combination of the potential terms which
obey the constraints.

\subsection{Spinorial cohomology at $\b=\ell^4$ }

By dimensional analysis $\S^{~~~a'}_{\a i}$ is, schematically,
\begin{align}
\S&\sim X^3\L\oplus X\nabla X\L \oplus X^2\nabla \L \oplus
X\nabla^2\L \oplus \nabla^2X\L
\oplus\nabla X\nabla \L \oplus\nabla^3\L \nn\\
&\oplus X^2\L^3 \oplus\nabla X\L^3 \oplus X\nabla\L\L^2
\oplus\L^2\nabla^2\L \oplus (\nabla\L)^2\L \oplus X\L^5
\oplus\nabla\L \L^4 \oplus \L^7 \label{cohfour}
\end{align}
In terms of irreducible representations of $Spin(1,2)\times
Spin(8)$, we can decompose the above contributions as
\begin{align}
\L^7
&\sim  7~(1)\times (1001)\oplus\dots\nn\\
\L^4\nabla\L
%&\sim
%\big( \yng(1,1,1,1)_{(3)}\times\yng(4)_{(0010)}\oplus
%\yng(4)_{(3)}\times\yng(1,1,1,1)_{(0010)}\oplus
%\yng(2,1,1)_{(3)}\times\yng(3,1)_{(0010)}\nn\\
%&\oplus \yng(3,1)_{(3)}\times\yng(2,1,1)_{(0010)}\oplus
%\yng(2,2)_{(3)}\times\yng(2,2)_{(0010)}\big)\otimes(5)\times(0010)\nn\\
&\sim 12~(1)\times(1001)\oplus \dots\nn\\
X^2\L^3
%&\sim
%\big(\yng(2)_{(4)}\times\yng(2)_{(1000)}\oplus\yng(1,1)_{(4)}
%\times\yng(1,1)_{(1000)}\big)\nn\\
%&\otimes\big(\yng(3)_{(3)}\times\yng(1,1,1)_{(0010)}
%\oplus\yng(2,1)_{(3)}
%\times\yng(2,1)_{(0010)}\big)
%\oplus\yng(1,1,1)_{(3)}
%\times\yng(3)_{(0010)}\big) \nn\\
&\sim 28~(1)\times (1001)\oplus\dots\nn\\
X^2\nabla\L
%&\sim \big(\yng(2)_{(4)}\times\yng(2)_{(1000)}\oplus\yng(1,1)_{(4)}
%\times\yng(1,1)_{(1000)}\big)
%\otimes
%(5)\times(0010)  \nn\\
&\sim 2~(1)\times (1001)\oplus\dots\nn\\
 \L^2\nabla^2\L
%&\sim \big(\yng(2)_{(3)}\times\yng(1,1)_{(0010)}\oplus\yng(1,1)_{(3)}
%\times\yng(2)_{(0010)}\big)
%\otimes
%(7)\times(0010)\nn\\
&\sim 1~(1)\times (1001)\oplus\dots\nn\\
 (\nabla\L)^2\L
%&\sim \big(\yng(2)_{(5)}\times\yng(1,1)_{(0010)}\oplus\yng(1,1)_{(5)}
%\times\yng(2)_{(0010)}\big)
%\otimes
%(3)\times(0010)\nn\\
&\sim 1~(1)\times (1001)\oplus\dots\nn\\
X\nabla X\L
%&\sim (4)\times(1000)\otimes
% (6)\times(1000)\otimes  (3)\times(0010)\nn\\
&\sim  4~(1)\times (1001)\oplus\dots\nn
\end{align}
In addition one can verify that there are no more contributions
coming from the  rest of the terms on the right-hand side of
(\ref{cohfour}). There are therefore fifty-five possible terms.

We shall now make the assumption that there is a non-trivial
cohomology element cubic in the fields. This is what we expect
from string theory calculations in ten dimensions. In order to
determine the cubic part $(\S^{(cub)a'}_{\a i})$ of the
cohomology, we shall only require the explicit form of the cubic
terms. These are eight in total,
\begin{align}
\S^{(1)a'}_{\a i}&=X^{b;ca'}\nabla_{b}X_{c}{}^{ab'}
(\s_{b'}\L_{a})_{\a i}\vert_{(1001)}  \nn\\
\S^{(2)a'}_{\a i}&=X^{bcb'}\nabla_{b}X_{c}{}^{aa'}
(\s_{b'}\L_{a})_{\a i}\vert_{(1001)} \nn\\
\S^{(3)a'}_{\a i}&=X^{bca'}\nabla_{b}X^{adb'}
(\c_{cd}\otimes \s_{b'}\L_{a})_{\a i}\vert_{(1001)}  \nn\\
\S^{(4)a'}_{\a i}&=X^{bcb'}\nabla_{b}X^{ada'}
(\c_{cd}\otimes \s_{b'}\L_{a})_{\a i}\vert_{(1001)} \nn\\
\S^{(5)a'}_{\a i}&=(\s_{b'}\nabla_b\L_c)_{\a i}
X^{bda'}X_{d}{}^{cb'}\vert_{(1001)}  \nn\\
\S^{(6)a'}_{\a i}&= (\c^{bc}\otimes\s_{b'}\nabla^d\L^e)_{\a i}
X_{bd}{}^{a'}X_{ce}{}^{b'}
\vert_{(1001)} \nn\\
\S^{(7)a'}_{\a i}&=(\L^b\c^d\otimes \s^{a'b'}\L^c)
(\s_{b'}\nabla_b\nabla_c \L_d)_{\a i}\vert_{(1001)} \nn\\
\S^{(8)a'}_{\a i}&=(\L^b\c^d\otimes\s^{a'b'}\nabla_b\L^c)
(\s_{b'}\nabla_c\L_d)_{\a i}\vert_{(1001)}  \nn, \label{upslns}
\end{align}
so that
\begin{equation}
\S^{(cub)a'}_{\a i}=\sum_{I=1}^{8}c^I\S^{(I)a'}_{\a i}.
\label{cubel}
\end{equation}
The coefficients $c^I$ will be determined, up to an overall
factor, in the following.

Carrying out a similar analysis for the redefinitions $(\d
z)^{a'}$ at this order in $\b$ we conclude that the only possible
contributions are of the form,
\begin{align}
\L^4X
%&\sim
%\big( \yng(1,1,1,1)_{(3)}\times\yng(4)_{(0010)}\oplus
%\yng(4)_{(3)}\times\yng(1,1,1,1)_{(0010)}\oplus
%\yng(2,1,1)_{(3)}\times\yng(3,1)_{(0010)}\nn\\
%&\oplus \yng(3,1)_{(3)}\times\yng(2,1,1)_{(0010)}\oplus
%\yng(2,2)_{(3)}\times\yng(2,2)_{(0010)}\big)\otimes(4)\times(1000)\nn\\
&\sim 5~(0)\times(1000)\oplus \dots\nn\\
\L^2\nabla X
%&\sim  \big(\yng(2)_{(3)}\times\yng(1,1)_{(0010)}\oplus\yng(1,1)_{(3)}
%\times\yng(2)_{(0010)}\big)\otimes
%(6)\times(1000)   \oplus \dots\nn\\
&\sim 1~(0)\times(1000) \oplus\dots\nn\\
X\L\nabla\L
%&\sim (4)\times(1000)\otimes (3)\times(0010) \otimes (5)\times (0010)\nn\\
&\sim 2~(0)\times(1100)\oplus \dots\nn\\
X^3&\sim 1~(0)\times(1100)\oplus \dots\nn
\end{align}
There are nine terms in total, but we shall only need the explicit
form of the following cubic terms,
\begin{align}
(\d z)^{(1)a'}&=X^{bca'}X_{bdb'}X_{c}{}^{db'}\nn\\
(\d z)^{(2)a'}&=(\L^b\c^d\otimes\s^{a'b'}\L^c)\nabla_bX_{cdb'}\nn\\
(\d z)^{(3)a'}&=(\L^b\c^d\otimes\s^{a'b'}\nabla_b\L^c)X_{cdb'}\nn\\
(\d z)^{(4)a'}&=(\L^d\c^b\nabla^c\L_d)X_{bc}{}^{a'}.
\label{redfns}
\end{align}
Finally, we need to repeat the analysis for the constraints
$Y_{a'b';c'}$ and $Y_{aa'b'c'd';e'}$. We find the following
contributions,
\begin{align}
X^2\nabla X&\sim 1~(0)\times(1100)\oplus \dots\nn\\
\nabla X\nabla \L\L&\sim 1~(0)\times(1100)\oplus \dots\nn\\
X\nabla^2\L\L&\sim 1~(0)\times(1100)\oplus\dots\nn\\
X^3\L^2&\sim 12~(0)\times(1100)\oplus 13~(2)\times(1002)\oplus\dots\nn\\
X\nabla\L\L^3&\sim 26~(0)\times(1100)\oplus 38~(2)\times(1002)\oplus\dots\nn\\
\nabla X\L^4&\sim 6~(0)\times(1100)\oplus 7~(2)\times(1002)\oplus\dots\nn\\
X\L^6&\sim 22~(0)\times(1100)\oplus
35~(2)\times(1002)\oplus\dots\nn
\end{align}
We shall only need the explicit form of the following terms,
\begin{align}
Y^{(1)a'b';c'}&=X^{bca'}X_a{}^{d[b'|}\nabla_bX^{ae|c']}
\vare_{cde}\vert_{(1100)}\nn\\
Y^{(1)a'b';c'}&=(\nabla^b\L^c\s^{b'c'}\L^a)\nabla_b
X_{ca}{}^{a'}\vert_{(1100)} \nn\\
Y^{(1)a'b';c'}&=(\nabla_b\nabla_c\L_a\s^{b'c'}\L^a)
X^{bca'}\vert_{(1100)}. \nn \label{cnstrts}
\end{align}
All the above is summarised in the following diagram.
%%
%%THE FIGURE
%%
%
\leavevmode
\begin{figure}[h]
\begin{center}
\input{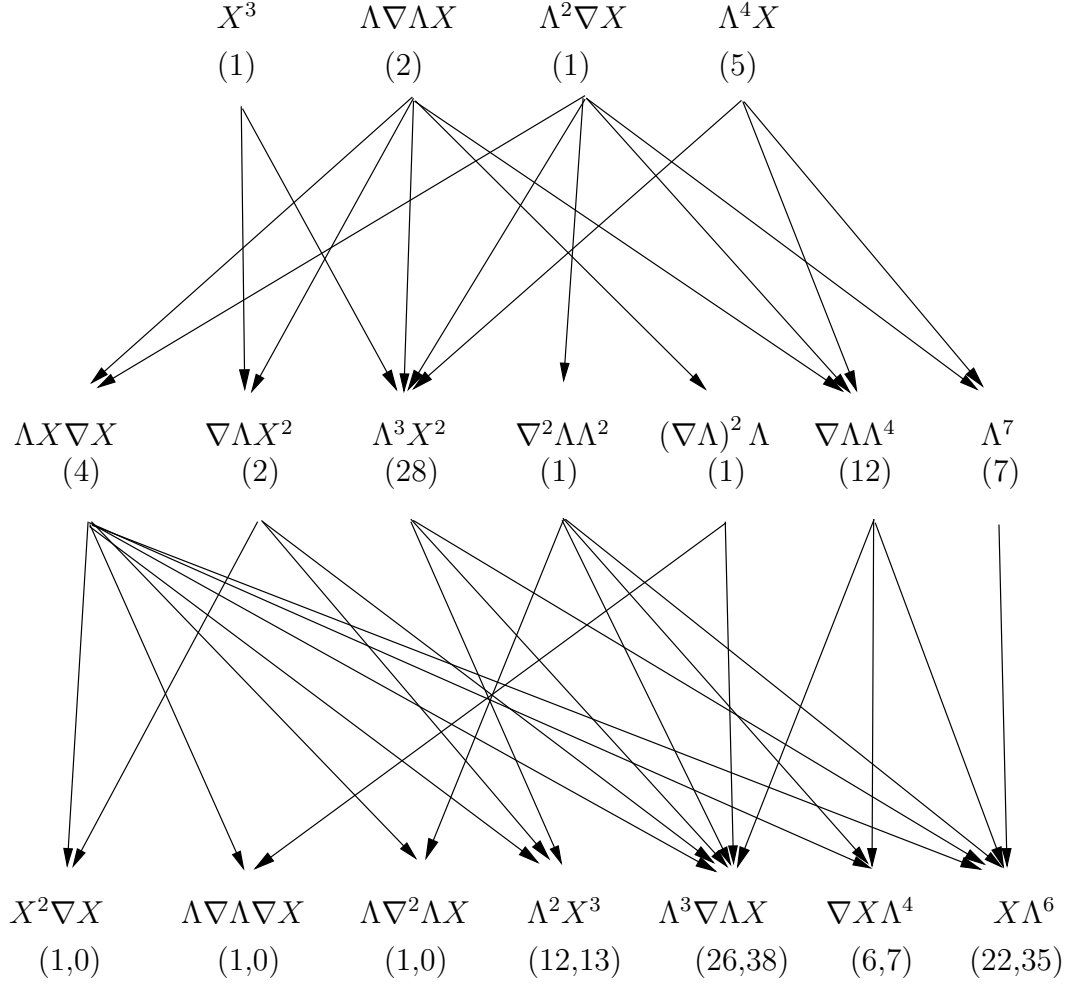}
\end{center}
\caption{Spinorial cohomology at $\ell^4$.}
\end{figure}
The coefficients $c^I$ in (\ref{cubel}) are determined, up to an
overall factor, as follows. First note that the redefinitions
(\ref{redfns}) can be used to eliminate $\S^{(I)a'}_{\a i};
~I=5\dots 8$. Indeed an explicit
 calculation gives
\begin{align}
d_s(\d z)^{(2)a'}&\to \S^{(7)a'}_{\a i}+\dots \nn\\
d_s(\d z)^{(3)a'}&\to i\S^{(8)a'}_{\a i}+\dots\nn,
\end{align}
so that $(\d z)^{(2,3)a'}$ can be used to eliminate
$\S^{(7,8)a'}_{\a i}$. Moreover,
\begin{align}
d_s(\d Z)^{(1)a'}&\to 2i\S^{(5)a'}_{\a i}+\dots \nn\\
d_s(\d Z)^{(4)a'}&\to-{1\over 2}\S^{(5)a'}_{\a i} +{1\over
2}\S^{(6)a'}_{\a i} +\dots \nn,
\end{align}
and so $(\d z)^{(1,4)a'}$ can be used to eliminate
$\S^{(5,6)a'}_{\a i}$. In the following we set
$$
c^I=0; ~~I=5\dots 8.
$$
Next we evaluate the action of $d_s$ on the remaining cubic terms,
\begin{align}
d_s\S^{(1)a'}_{\a i}&\to-4iY^{(3)a'b';c'}  +\dots \nn\\
d_s\S^{(2)a'}_{\a i}&\to-4iY^{(2)a'b';c'}  +\dots \nn\\
d_s\S^{(3)a'}_{\a i}&\to-16iY^{(1)a'b';c'}-4iY^{(3)a'b';c'}  +\dots  \nn\\
d_s\S^{(4)a'}_{\a i}&\to-32iY^{(1)a'b';c'}+4iY^{(2)a'b';c'}
+\dots \nn,
\end{align}
where the ellipses stand for terms with more than three fields.
Demanding the vanishing of $d_s$ fixes the remaining coefficients
to be
\begin{align}
c^2&={1\over 2}c^1\nn\\
c^3&=-c^1\nn\\
c^4&={1\over 2}c^1\nn.
\end{align}
We conclude that the part of the spinorial cohomology cubic in the
fields at order $\ell^4$ is determined up to an overall factor to
be represented by

 \bea
 \S^{~~~a'}_{\a i}&=&X^{bca'}\nabla_{b}X_{c}{}^{ab'}
(\s_{b'}\L_{a})_{\a i}\vert_{(1001)}+ \frac{1}{2}
X^{bcb'}\nabla_{b}X_{c}{}^{aa'}
(\s_{b'}\L_{a})_{\a i}\vert_{(1001)} \nn\\
&&-X^{bca'}\nabla_{b}X^{adb'} (\c_{cd}\otimes \s_{b'}\L_{a})_{\a
i}\vert_{(1001)} \nn\\
&&+\frac{1}{2} X^{bcb'}\nabla_{b}X^{ada'} (\c_{cd}\otimes
\s_{b'}\L_{a})_{\a i}\vert_{(1001)}
 \eea

\section{Action}

There are various ways of deriving the action corresponding to the
deformed embedding constraint. One is to compute the modified
equations of motion by solving the torsion constraints at
dimensions one-half and one, and then to work back to the
Lagrangian. This is a systematic approach, but is fairly tedious
in practice. Another possibility is to use the action principle
for branes. We shall discuss this below, but it turns out that one
is required to solve another spinorial cohomology problem. In
order to avoid this we shall find the key part of the Lagrangian
up to quadratic order in fermions by brutal force. Nevertheless,
the action principle does shed light on the structure of the
component action.

\subsection{Action principle}

We recall that  supersymmetric Lagrangians in $D$-dimensional
spacetime can be constructed from closed superspace $D$-forms
\cite{D'Auria:1982pm,ggks}. Given such a form, $L$, the spacetime
action is

 \be
 S=\int\,\e^{m_1\ldots m_D} L_{m_1\ldots m_D}(x,0)
 \la{act}
 \ee

Under an infinitesimal superspace diffeomorphism generated by a
vector field $v$, $\d L=\cL_v L=i_v dL+ d(i_v L)=d(i_v L)$ because
$L$ is closed. By evaluating this variation at $\th=0$ on the
spacetime part of the $D$-form one can see that the action
\eq{act} is invariant under spacetime diffeomorphisms and local
supersymmetry transformations, these being identified as the
leading components of the even and odd superspace diffeomorphisms.

In the context of branes one can construct a Lagrangian form
starting from the (closed) Wess-Zumino $(d+1)$-form $W$ on the
super worldvolume, where $d$ is the dimension of the bosonic
worldvolume. This can be written in terms of an explicit potential
$Z$, $W=d Z$, but it can also be written in terms of a
globally-defined $d$-form, $W=dK$. The $d$-form $L=K-Z$ is closed
by construction. It is uniquely defined up to the exterior
derivative of a $(d-1)$-form which is irrelevant in the action.
When the standard embedding constraint $E_{\a}{}^{\una}=0$ is
imposed (in the on-shell or Lagrangian off-shell cases) this
procedure gives the Green-Schwarz action, including the
Born-Infeld contribution in the case of $D$-branes \cite{hrs}. At
higher orders, it must still be possible to write $W$ as $dK$, but
we shall also require Lagrangian forms satisfying $dL=0$.

For the M2-brane $W$ is the pull-back of the target space
four-form $H$ on to the brane, $W=f^*\unH:=H$. The components of
the pullback are (up to signs)
$$
H_{ABCD}=E_D{}^{\unD}E_C{}^{\unC}
E_B{}^{\unB}E_A{}^{\unA}H_{\unA\unB\unC\unD}.
$$
The only non-vanishing component of $\unH$ in flat superspace is

 \be
 H_{\una\unb,\uc\ud}=-i(\C_{\una\unb})_{\uc\ud}
 \ee

To order $\b$ we therefore have

\noindent dim. -1
\begin{equation}
H_{\a\b\c\d}=0 \label{ha}
\end{equation}
\noindent dim. -1/2
\begin{equation}
H_{a\b\c\d}=0
\end{equation}
\noindent dim. 0
\begin{equation}
H_{ab\c\d}=-i(\C_{ab})_{\c\d}-4i\L_{[a}{}^{\a'}(\C_{b]c'})_{\a'(\c}
\psi_{\d)}{}^{c'}
\end{equation}
\noindent dim. 1/2
\begin{equation}
H_{abc\d}=-3i(\C_{[ab|})_{\c'\d'}\L_{|c]}{}^{\c'}h_{\d}{}^{\d'}
\end{equation}
\noindent dim. 1
\begin{equation}
H_{abcd}=0 \label{hd}
\end{equation}
We can now solve $H=dK$. In components,
$$
H_{ABCD}=4\nab_{[A}K_{BCD]}+ 6 T_{[AB|}^FK_{F|CD]}
$$
Note that for $\b=0$, all the components of $K$ are zero, except
for the highest-dimension component $K_{abc}$. To order $\b$ we
have, taking (\ref{ha}--\ref{hd}) into account ,

\noindent dim. -1
$$
0=-{3i\over 2}(\C^f)_{(\a\b|}K_{f|\c\d)}+\nab_{(\a}K_{\b\c\d)}
$$
This is solved by
$$
K_{\a\b\c}=0; ~~~~~K_{a\b\c}=0.
$$

\noindent dim. -1/2
$$
0=(\C^f)_{(\b\c|}K_{fa|\d)}
$$
This implies
$$
K_{ab\c}=0.
$$
\noindent dim. 0
$$
-i(\C_{ab})_{\c\d}-4i\L_{[a}{}^{\a'}(\C_{b]c'})_{\a'(\c}
\psi_{\d)}{}^{c'}=
\big(-i(\C^e)_{\c\d}-2i\L^{e\a'}(\C_{c'})_{\a'(\c}
\psi_{\d)}{}^{c'} \big)K_{eab}
$$
where we have taken into account the expression for $T_{\a\b}{}^e$
to order $\b$, given in section three.  The above equation is
satisfied when
$$
K_{abc}=\vare_{abc}.
$$
This is identical to the solution for $K_{ABC}$ in the $\b=0$
case. This does not necessarily mean that there is no correction
to the Green-Schwarz action because the induced Green-Schwarz
metric may change. However, one can check that there is no such
change to first order in $\b$ although corrections of this type
will arise at higher orders.

This means that the Lagrangian form we are looking for must
satisfy $dL=0$, and, furthermore, that $L_{\a\b\c}$ cannot be
zero. Since the Lagrangian form can be changed by an exact form
without affecting the action \eq{act}, it follows that this gives
rise to a new spinorial cohomology problem. This is more
complicated than the one treated in section six in that the
representations involved are larger. On dimensional grounds, the
order $\b$ contribution to $L_{\a\b\c}$ which is quartic in fields
has the schematic form $\L^3 X$ but we have not attempted to
compute it explicitly. However, it is useful to observe that this
method of constructing actions tells us something about the
structure of the component Lagrangian. From \eq{act} we have

 \be
 \e^{mnp} L_{mnp}=\e^{mnp}\left(E_p{}^c E_n{}^b E_m{}^a L_{abc} +
 3E_p{}^c E_n{}^b E_m{}^{\a} L_{\a bc}+\ldots\right)
 \la{lag}
 \ee

The non-leading terms on the right-hand side all involve the
induced worldvolume ``gravitino'' field $E_m{}^{\a}$. This is
related to  $\del_m\th^{\ua}$ but it is not expressible in terms
of $\L$. The components $L_{ABC}$ will all be constructed from
manifest tensors, namely $\L$ and $X$; the gravitino terms reflect
the fact that the Lagrangian is a density.

\subsection{Component approach}

In the component approach we write the action as $S=S_0+S_1$,
where $S_0$ is the usual Green-Schwarz action for the
supermembrane \cite{Bergshoeff:1987cm}, and similarly for the
worldvolume supersymmetry variations, so that at first order we
have

\be \d_1 S_0 + \d_0 S_1 = 0 \ee

We can ignore the first term if we calculate $S_1$ using the
lowest-order field equations. The neglected terms correspond to
field redefinitions. We are therefore looking for an on-shell
invariant at order $\b$.

{}From the discussion above we have seen that the first-order
Lagrangian density $\cL$ will have the form

 \be
 \cL=\sqrt{-g}\hat L + \e^{mnp}\left(
 3E_p{}^c E_n{}^b E_m{}^{\a} L_{\a bc}+\ldots\right)
 \la{82}
 \ee

where $g$ is the determinant of the Green-Schwarz metric, $\hat
 L:= \e^{abc} L_{abc}$ and where the right-hand side is understood
to be evaluated at $\th=0$. Under a worldvolume supersymmetry
variation we find

 \bea
 \d\cL &=& \del_m (v^{\a} \cL^m_{\a})\nn\\
 &=& \d(\sqrt{-g})\hat L + \sqrt{-g}\, \d \hat L \nn\\
 &\phantom{=}& +\e^{mnp}\left(3E_p{}^c E_n{}^b E_m{}^{\a}\d L_{\a bc}
 +6E_p{}^c (\d E_n{}^b) E_m{}^{\a}L_{\a
 bc}+3E_p{}^c  E_n{}^b (\d E_m{}^{\a})L_{\a
 bc} \ldots\right)
 \eea

for some $\cL^m_{\a}$.  To quadratic order in fermions the terms
represented by the dots in \eq{82}, which have two or more powers
of $E_m{}^{\a}$, can be disregarded. The schematic form of the
Lagrangian terms are

 \bea
 \hat L&\sim& X^4 + \L^2 \nab X^2 + \cO(\L^4) \nn\\
 L_{\a bc}&\sim& \L X^3 + \cO(\L^3)
 \eea

while

 \bea
 \d E_m{}^a &=&iv^{\b}(\C^a)_{\b\c} E_m{}^{\c}\nn\\
 \d E_m{}^{\a}&=& \nab_m v^{\a} + v^{\b}E_m{}^c T_{\b c}{}^{\a}~.
 \eea

We shall focus on the terms in the variation of $\cL$ which are
linear in fermions. In the variation of $E_m{}^{\a}$ we can use
the lowest-order expression for the induced torsion $T_{\b
c}{}^{\a}$. As this is quadratic in $\L$ the corresponding term in
$\d E_m{}^\a$ can be ignored. The term involving the variation of
the world-volume vielbein $E_m{}^a$ can also be ignored as it
leads to terms which are cubic in  fermions. The terms which are
linear in fermions involve either $\L$ or $E_m{}^{\a}$, and these
must cancel separately. Thus we see that the term arising from the
variation of the determinant of the metric multiplied by the $X^4$
term in $\hat L$ must be cancelled by the leading term coming from
the variation of $L_{\a bc}$. The terms linear in $\L$ arise from
varying both the $\L^0$ and $\L^2$ terms in $\hat L$ and from the
$\nab v$ variation term in $E_m{}^{\a}$ multiplied by $L_{\a bc}$.
It therefore follows that the terms linear in $\L$ in the
variation of $\hat L$ will cancel up to a covariant divergence if
the parameter $v^{\a}$ is taken to be covariantly constant. The
terms involving the derivative of $v$ in the variation of $\hat L$
must cancel against similar terms arising from the variation of
$E_m{}^{\a}$. In fact, demanding that these terms cancel
determines $L_{\a bc}$ at first order in $\L$. We shall therefore
concentrate on computing $\hat L$ up to quadratic order in the
fermions. In what follows we shall refer to $\hat L$ as the
Lagrangian.

It follows from the analysis in section six that, up to and
including terms quadratic in the fermions, there is a unique
supersymmetric invariant at order $l^4$. Indeed, we have shown
that at order cubic in the fields there is a unique nontrivial
element in the relevant spinorial cohomology group and this
implies that there is a unique deformation of the equations of
motion. We conclude that in the action there is a unique
supersymmetric invariant at order quartic in the fields. In
particular,  there can be no terms in the action of order cubic,
or less, in the fields. There are also no terms of the form
$\Lambda^2X^3$ as can be seen by the fact that there is no scalar
in the decomposition of the tensor product
\be
[(3)\times(0010)]^{2\otimes_a}\otimes[(4)\times(1000)]^{3\otimes_s},
\ee
The bosonic part of the Lagrangian at $l^4$, is a linear
combination of the following three $X^4$ terms
\bea
t_1&:=&X_{ab}{}^{a'}X_{cda'}X^{abb'}X^{cd}{}_{b'} \nn\\
t_2&:=&X_{ab}{}^{a'}X_{cd}{}^{b'}X^{ab}{}_{a'}X^{cd}{}_{b'} \nn\\
t_3&:=&X_{ab}{}^{a'}X_{cd}{}^{b'}X^{ac}{}_{a'}X^{bd}{}_{b'}. \eea
A fourth structure
\be t_4:=X_{ab}{}^{a'}X_{cda'}X^{acb'}X^{bd}{}_{b'} \ee
can be shown to be linearly dependent
\be t_4=t_1+{1\over 2}t_2-2t_3, \label{wittgenstein} \ee
by using the identity
\be 4X_{[a}{}^{[b|a'}   X_{c]}{}^{|d]b'}= \left(\delta_f^e
X^{gha'}X_{gh}{}^{b'}
-X_f{}^{ga'}X_g{}^{eb'}-X_f{}^{gb'}X_g{}^{ea'}  \right)
\varepsilon_{eac} \varepsilon^{fbd}. \label{rc} \ee
At quadratic order in the fermions, the Lagrangian at $l^4$
contains terms of the form $\Lambda^2X\nabla X$ and $\Lambda
\nabla\Lambda X^2$. A lengthy calculation, the details of which
are given in appendix B, yields the following result for $\hat L$:

\bea \hat L&=&{1\over 4}X_{ab}{}^{a'}X_{cda'}X^{abb'}X^{cd}{}_{b'}
-{1\over 4}X_{ab}{}^{a'}X_{cd}{}^{b'}X^{ac}{}_{a'}X^{bd}{}_{b'}\nn\\
&&+i(\Lambda_a\gamma_b\nabla_c\Lambda_d) X^{aba'}X^{cd}{}_{a'}
-{2i\over
3}(\Lambda^f\gamma_a\otimes\sigma^{a'b'}\nabla_f\Lambda_b)
X^{ae}{}_{a'}X^{b}{}_{eb'}\nn\\
&&-{i\over
3}(\Lambda_a\gamma_b\otimes\sigma^{a'b'}\nabla_c\Lambda_d)
X^{ab}{}_{a'}X^{cd}{}_{b'}+{\cal O}(\Lambda^4). \label{alle} \eea
The Lagrangian above is determined up to a total derivative and up
to terms vanishing by virtue of the lowest-order equations of
motion. The latter can always by removed by appropriate field
redefinitions.

We shall now compare our result to that of \cite{Bachas:1999um}.
In this paper the bosonic CP-even part of the ${\cal O}(l^4)$
Lagrangian of a D-$p$ brane was determined to be
\be {1\over 16}\left( (R_T)_{abcd}(R_T)^{abcd}
-2(R_T)_{ab}(R_T)^{ab}-(R_N)_{abc'd'}(R_N)^{abc'd'}
+2\bar{R}_{a'b'}\bar{R}^{a'b'} \right), \label{bachas} \ee
see equation (2.14) of that reference. In the above we have
omitted an overall multiplicative constant involving $\alpha'$ and
the string coupling.

Let us view (\ref{bachas}) as the world-volume Lagrangian of a
D2-brane and consider its lift to eleven dimensions. The various
curvatures in (\ref{bachas}) are given, in the notation of the
present paper and for flat target-space, by
\bea
(R_T)_{abcd}&=&X_{aca'}X_{bd}{}^{a'}-X_{ada'}X_{bc}{}^{a'}   \nn\\
(R_N)_{abc'd'}&=&X_{aec'}X^e{}_{bd'}-X_{bec'}X^e{}_{ad'} \nn\\
(R_T)_{ab}&=&-X_{ae}{}^{a'}X^e{}_{ba'}\nn\\
\bar{R}_{a'b'}&=&X_{aba'}X^{ab}{}_{b'}. \label{plop} \eea
Substituting (\ref{plop}) into (\ref{bachas}) on gets precisely
the bosonic part of (\ref{alle}).

The Lagrangian (\ref{bachas}) was determined, as is the case with
(\ref{alle}), modulo total derivatives \footnote{the coefficient
of the Gauss-Bonnet term was fixed to zero in \cite{Bachas:1999um}
by an indirect argument.} and terms which vanish by virtue of the
lowest-order equations of motion. In \cite{Fotopoulos:2001pt} the
investigation of these ambiguities was considered, see equation
(5.4) therein. It was shown that there is a total of five
ambiguous terms, two of which vanish by virtue of the lowest-order
equations of motion and therefore cannot be determined
perturbatively. Two of the other three terms vanish for flat
target-space, whereas the remaining third term vanishes by virtue
of equation (\ref{wittgenstein}) of the present paper.

\section{Kappa symmetry}

Although we have not used the modified worldvolume supersymmetry
transformations in determining the Lagrangian, it is perhaps of
interest to see how the usual $\k$-symmetry variations are
altered. To lowest order the $\k$-symmetry transformations were
derived, starting from worldvolume supersymmetry transformations,
in equations (9-14). These will be amended when we deform the
embedding condition. In general, a worldvolume supersymmetry
transformation gives the following transformations

 \bea
 \d z^{\una}&=& v^\a E_{\a}{}^{\una}  \\
 \d z^{\ua}&=& v^{\a} E_{\a}{}^{\ua}~.
 \eea

We can convert the second of these into $\k$-symmetry form as
before:

 \be
 \d z^{\ua}=\k^{\ub}
 P_{\ub}{}^{\ua}=\frac{1}{2}\k^{\ub}(1+\C)_{\ub}{}^{\ua}~.
 \ee

where $\k^{\ua}=v^{\a} E_{\a}{}^{\ua}$ and

 \be
 P_{\ua}{}^{\ub}=(E^{-1})_{\ua}{}^{\c} E_{\c}{}^{\ub}
 \ee

where the inverse refers to inverting the square matrix formed
from $(E_{\a}{}^{\ua},E_{\a'}{}^{\ua})$. Explicitly

 \be
 P_{\ua}{}^{\ub}=(u^{-1})_{\ua}{}^{\c}u_{\c}{}^{\ub} +
 (u^{-1})_{\ua}{}^{\c}h_{\c}{}^{\d'}u_{\d'}{}^{\ub}
 \ee

The first term here has the same structure as in the undeformed
theory. The bosonic projection of $\d z^{\unM}$ can be written

 \be
 \d z^{\una}=\k^{\ua} \Psi_{\ua}{}^{\una}=\d z^{\ua}
 \Psi_{\ua}{}^{\una}~,
 \ee

where

 \bea
 \Psi_{\ua}{}^{\una}&=&(E^{-1})_{\ua}{}^\c E_{\c}{}^{\una}\nn\\
 &=&(E^{-1})_{\ua}{}^\c \psi_{\c}{}^{a'} u_{a'}{}^{\una}
 \eea

satisfies $\Psi=P\Psi$.

In the undeformed theory the matrix $\C$ is given by

 \be
 \C=-\frac{1}{6}\e^{abc} u_a{}^{\una} u_b{}^{\unb} u_c{}^{\unc}
 \C_{\una\unb\unc}~.
 \la{plato}
 \ee

This can easily be seen to be the same as

 \be
 \C_{\ua}{}^{\ub}=(u^{-1})_{\ua}{}^{\c}u_{\c}{}^{\ub}-
 (u^{-1})_{\ua}{}^{\c'}u_{\c'}{}^{\ub}
 \la{socrates}
 \ee

In the deformed theory we can write

 \be
 \C=\C_0 +\frac{1}{2}u^{-1}(1+\C_0)h(1-\C_0)u
 \ee

where $\C_0$ has the same form  as in \eq{plato} and \eq{socrates}
and where $h$ is regarded as a $32\xz 32$ matrix satisfying $h=P_0
h Q_0$, with $P_0=1/2(1+\C_0),\ Q=1/2(1-\C_0)$.

The formulae above are valid for an arbitrary deformation $\psi$.
To obtain the order $\b$ term we merely have to substitute into
the expressions the explicit forms for $\psi$ and $h$.

\section{Conclusions}

In this paper we have discussed the first correction to the
dynamics of the M2-brane in the superembedding formalism which we
have shown to occur at order $\ell^4$. We have seen that the
allowed modifications of the constraints are specified by an
element of a certain spinorial cohomology group. There is a unique
solution cubic in the fields and we believe that there are no
other independent terms at quartic or higher orders. In principle
one could continue this analysis for higher powers of $\ell$. We
would expect to find both corrections induced by iterating the
first correction and new cohomological terms. However, the
computations are difficult, even at $\ell^4$.

The formalism developed here could be applied straightforwardly to
other branes such as the M5-brane and D-branes. Indeed, we could
obtain the equivalent terms in the D2-brane action by dimensional
reduction. In particular, this would shed some light on the
$\k$-symmetric effective action including $\del^4 F^4$ terms.
Another extension of the formalism would be to include non-trivial
supergravity backgrounds. In $D=11$ it is  expected that the first
correction to supergravity occurs at $\ell^6$, so that this
complication could be ignored. It should therefore be relatively
straightforward to compute the corrections at the same order,
which involve the background supergravity fields. In the case of
the M5-brane these results could again be dimensionally reduced to
ten dimensions, either to give results for the NS5-brane, or the
D4-brane. In the latter case one would expect to see some sign of
the $\hat A$-genus terms discussed in
\cite{Green:1996dd,Cheung:1997az,minmo}. In principle, we should
therefore be able to obtain some information about the
supersymmetrisation of these terms.

\section*{Acknowledgements}

This work was supported in part by EU contracts HPRN-2000-00122 
(which includes Queen 
Mary, London as a subcontractor)
and HPRN-CT-2000-00148 and PPARC grants PPA/G/S/ 1998/00613 and 
PPA/G/O/2000/00451. 
SFK thanks the German National Merit Foundation for
financial support. UL acknowledges support in part by VR grant
650-1998368.

%%%%%%%%%%%%   APPENDIX    %%%%%%%%%%%%%%%

\section*{Appendix A}

\subsection*{Notation}

The following index conventions are used: plain (underlined)
indices refer to worldsurface and target space quantities
respectively and primed indices refer to normal spaces; indices
from the beginning of the alphabet refer to preferred bases,
indices from the middle of the alphabet to coordinate bases; Latin
(Greek) indices are used for even (odd) indices, while capital
indices run over the whole space. Thus a worldsurface preferred
basis index could be $A=(a,\a)$ while a target space coordinate
index could be $\unM=(\unm,\um)$. Tensor quantities with indices
are not underlined, but if the indices are omitted or if a target
space tensor is projected on some of its indices we underline the
tensor.

For the spinor indices a two step notation is used. Initially the
target space index $\ua$, running from 1 to 32 is split in two,
$\ua\rightarrow(\a,\a')$, where both the worldsurface index $\a$
and the normal index $\a'$ run from 1 to 16. For some purposes
this is adequate but sometimes one wishes to recognise explicitly
that these indices transform under $Spin(1,2)\xz Spin(8)$. We set

\bea
\psi^{\a}&\rightarrow&  \psi^{\a i}\nonumber\\
\psi^{\a'}&\rightarrow& \psi^{\a i'} \eea where the spinor index
on the right takes on two values while $i$ and $i'$ both  run from
1 to 8. We use space-favoured metrics throughout,
$\h_{ab}=(-1,+1,\ldots,+1)$, and the spacetime $\e$-tensors have
$\e^{0123\ldots}=+1$.

We use the following representation of the $D=11$ $\C$-matrices:
\bea
\C^a&=&\c^a\otimes \c_9\nonumber \\
\C^{a'}&=& 1\otimes \c^{a'} \eea where the $d=3$ $\c$-matrices are
$2\xz 2$ and the $d'=8$ $\c$-matrices are $16\xz 16$. The charge
conjugation matrix is \be C=\e\otimes\c_9 \ee In indices \bea
(\C^a)_{\ua}{}^{\ub}&=&(\c^a)_{\a}{}^{\b}\left(\ba{cc} \d_i{}^j & 0\\
0 & -\d_{i'}{}^{j'}\ea\right) \nonumber\\
(\C^{a'})_{\ua}{}^{\ub}&=& \d_{\a}{}^{\b}\left(\ba{cc} 0&
(\s^{a'})_{i}{}^{j'}\\
(\tilde\s^{a'})_{i'}{}^{j}&0\ea\right) \eea and \be
C_{\ua\ub}=\e_{\a\b}\left(\ba{cc}\d_{ij} &0\\
0&-\d_{i'j'}\ea\right) \ee where the $8\xz 8$ $\s$-matrices are
related to the eight-dimensional $\c$-matrices by \be
\c^{a'}=\left(\ba{cc}0&(\s^{a'})_{ij'}\\
(\tilde\s^{\a'})_{i'j} &0\ea\right) \ee Eleven-dimensional spinor
indices are raised or lowered according to the rule \be
\psi^{\ua}=C^{\ua\ub} \psi_{\ub}\ \leftrightarrow\
\psi_{\ua}=\psi^{\ub}C_{\ub\ua} \ee where $C$ with upper indices
is the same matrix as $C$ with lower indices. Three-dimensional
spinor indices are similarly raised or lowered using $\e$.
Eight-dimensional indices, whether spinor or vector, are raised
and lowered with the standard Euclidean metric. The $\C$-matrices
with lowered indices are \bea
(\C^a)_{\ua\ub}&=&(\c^a)_{\a\b}\left(\ba{cc} \d_{ij} & 0\\
0 & \d_{i'j'}\ea\right) \nonumber\\
(\C^{a'})_{\ua\ub}&=& \e_{\a\b}\left(\ba{cc} 0&-(\s^{a'})_{ij'}\\
(\tilde\s^{a'})_{i'j}&0\ea\right) \eea It is straightforward to
decompose any of the eleven-dimensional $\C$-matrices with
multi-vector indices in this way. We give the two-index
$\C$-matrices as they are used most in the text: \bea
(\C^{ab})_{\ua\ub}&=&(\c^{ab})_{\a\b}\left(\ba{cc} \d_{ij} & 0\\
0 & \d_{i'j'}\ea\right) \nonumber\\
(\C^{ab'})_{\ua\ub}&=&-(\c^a)_{\a\b}\left(\ba{cc} 0&-(\s^{b'})_{ij'}\\
(\tilde\s^{b'})_{i'j}&0\ea\right)\nonumber\\
(\C^{a'b'})_{\ua\ub}&=& \e_{\a\b}\left(\ba{cc} (\s^{a'b'})_{ij'}&0\\
0&-(\tilde\s^{a'b'})_{i'j}\ea\right) \eea

The three-dimensional $\c$-matrices are real, symmetric with
lowered indices, and satisfy \be \c^a\c^b=\h^{ab} + \c^{ab} \ee
where \be \c^{ab}=\e^{abc}\c_c\ \leftrightarrow
\c_a=-{1\over2}\e_{abc}\c^{bc} \ee

For the eight-dimensional $\s$-matrices one may take \bea
\s^{a'}&=&(1,i\t_r)\nonumber\\
\tilde\s^{a'}&=&(1,-i\t_r) \eea where $r=1,\ldots 7$, and where
the $\t_r$ are seven-dimensional Dirac matrices which are purely
imaginary and antisymmetric. The matrices $\s^{a'b'c'd'}$ and
$\tilde\s^{a'b'c'd'}$ are symmetric, the former being self-dual,
the latter anti-self-dual, $\s^{a'b'}$ and $\tilde\s^{a'b'}$ are
antisymmetric.

\subsection*{Group theoretic conventions}

Consider a Lie group $G$ and a $G$-module $V$. The $n$-th tensor
product $V^{\otimes n}$ admits a decomposition
$$
\sum_{R} V_R \times R
$$
under $G\times S_n$, where $R$ runs over all irreducible
representations of the symmetric group $S_n$ and $V_R$ is a
$G$-module. As is well known, the irreducible representations of
$S_n$ can be parametrised by partitions of $n$ or, equivalently,
by the associated Young diagrams. If $R$ is associated to the
partition $\l$ of $n$, $V_R$ is the {\sl plethysm} of $V$ with
respect to $\l$.

In this paper, we are using a notation referring to the product
group $Spin(1,2)\times Spin(8)$.

The highest weight of the standard module (i.e. of the vector
representation) of $Spin(8)$ is given by $(1000)$ on the basis of
fundamental weights. A two-form is $(0100)$ and a three-form is
$(0011)$. The self-dual and anti-self-dual four-forms are
represented by $(0002)$, $(0020)$ respectively. The chiral and
anti-chiral spinors are $(0001)$, $(0010)$. Moreover, the highest
weight of the $k$-form-spinor is given by the sum of the highest
weight of the $k$-form and the highest weight of the spinor.
Recall that by a $k$-form-spinor we refer to the projection onto
the irreducible (gamma-traceless) part. For example $(1001)$ is a
chiral vector-spinor, $(0110)$ is an anti-chiral two-form-spinor,
etc. We can also consider the projection onto the highest-weight
representation of the plethysm associated to the partition
$[2,\dots  ,2,1,\dots ,1]$ of $n+k$, with $k$ entries equal to 2
and $n-k$ entries equal to 1. This is one way to view the
irreducible $(n,k)$-tensors of section 5. For example, a
(2,1)-tensor is a two-form and a one-form and is represented by
$(1100)$. A (3,1)-tensor is a three-form and a one-form and is
represented by $(1011)$, etc. The highest-weights of irreducible
(gamma-traceless) $(n,k)$-tensor-spinors are represented in the
obvious way. For example $(1101)$ is the highest weight of a
$(2,1)$-tensor-(chiral) spinor. This discussion generalizes
straightforwardly to representations corresponding to more general
partitions.

In the case of $Spin(1,2)$ the situation is simpler, in that the
representation denoted by $(n)$ is of dimension $n+1$. Hence, the
spinor representation is denoted by $(1)$, the vector by $(2)$,
the vector spinor by $(3)$ and so on.

%%%%%%%%%%%APPENDIX   B%%%%%%%%%%%%%%%%%%%%%%%%%%%%%%%%

\section*{Appendix B}

In this appendix we give some details of the derivation of the
action in section seven. At quadratic order in the fermions the
Lagrangian at $\ell^4$ contains terms of the form $\L^2 X\nab X$
and $\L\nab\L X^2$. We have the following structures:

\noindent $\Lambda^2X\nabla X$:
\bea u_1&:=&(\Lambda^f\gamma_a\otimes\sigma^{a'b'}\Lambda_f)
X_{cda'}\nabla^aX^{cd}{}_{b'}\nn\\
u_2&:=&(\Lambda_a\gamma_b\otimes\sigma^{a'b'}\Lambda_c)
X^{a}{}_{fa'}\nabla^fX^{bc}{}_{b'}\nn\\
u_3&:=&(\Lambda_a\gamma_b\Lambda_c)
X^{a}{}_{fa'}\nabla^fX^{bca'}\nn\\
u_4&:=&(\Lambda_a\gamma_b\otimes\sigma^{a'b'}\Lambda_c)
X^{b}{}_{fa'}\nabla^fX^{ac}{}_{b'} \eea
\noindent $\Lambda \nabla\Lambda X^2$:
\bea s_1&:=&(\Lambda^f\gamma_a\nabla_f\Lambda_b)
X^{aea'}X^{b}{}_{ea'}\nn\\
s_2&:=&(\Lambda_a\gamma_b\nabla_c\Lambda_d)
X^{aba'}X^{cd}{}_{a'}\nn\\
s_3&:=&(\Lambda^f\gamma_a\otimes\sigma^{a'b'}\nabla_f\Lambda_b)
X^{ae}{}_{a'}X^{b}{}_{eb'}\nn\\
s_4&:=&(\Lambda_a\gamma_b\otimes\sigma^{a'b'}\nabla_c\Lambda_d)
X^{ab}{}_{a'}X^{cd}{}_{b'}\nn\\
s_5&:=&(\Lambda_a\gamma_b\otimes\sigma^{a'b'}\nabla_c\Lambda_d)
X^{ac}{}_{a'}X^{bd}{}_{b'}\nn\\
s_6&:=&(\Lambda_a\gamma_b\nabla_c\Lambda_d) X^{aca'}X^{bd}{}_{a'}.
\eea
Not all of the above are independent. Only three of the $u$ and
four of the $s$-terms are, as can be seen by counting the number
of scalars in the decomposition of the tensor products
\be [(3)\times(0010)]^{2\otimes_a}\otimes[(4)\times(1000)]
\otimes[(6)\times(1000)] \ee
and
\be [(3)\times(0010)]\otimes [(5)\times(0010)]
\otimes[(4)\times(1000)]^{2\otimes_s}. \ee
Indeed on has
\bea
0&=&u_2-u_4-{1\over 2}u_1\nn\\
0&=&s_6-s_2+s_1\nn\\
0&=&3s_5-s_4+s_3. \label{ra} \eea
The first line of (\ref{ra}) is derived using the identity
\be
\varepsilon_{egh}(\Lambda^g\gamma^h\otimes\sigma^{a'b'}\Lambda_c)
={1\over 2} \varepsilon_{dec}
(\Lambda^f\gamma^d\otimes\sigma^{a'b'}\Lambda_f). \label{rb} \ee
The second line of (\ref{ra}) can be shown by taking (\ref{rc})
into account.
Finally, to derive the third line of  (\ref{ra}) one notes that
\be
0=(\Lambda_e\gamma^{[ae}\gamma_b\otimes\sigma_{a'b'}\nabla_c\Lambda_d)
X_{a}{}^{c|a'}X^{|b]db'}. \label{rd} \ee
In addition one has to take into account the fact that some terms
are related to each other by integration by parts. Indeed, up to
total derivatives, one has the following relations
\bea
u_1&=&2s_3\nn\\
u_2&=&s_4-s_5\nn\\
u_2&=&u_4+s_3\nn\\
u_3&=&s_2-s_6\nn\\
u_3&=&s_1\nn\\
u_4&=&2s_5\nn\\
u_4&=&u_2-s_3. \label{re} \eea
Only six of the equations (\ref{ra},\ref{re}) above are
independent. It is perhaps interesting to note that the
group-theoretic relations (\ref{ra}) are all implied by the
partial integration equations (\ref{re}). The solution may be
given in terms of four independent variables which we take to be
$s_{1-4}$,
\bea
u_1&=&2s_3\nn\\
u_2&=&{1\over 3}s_3+{2\over 3}s_4\nn\\
u_3&=&s_1\nn\\
u_4&=&-{2\over 3}s_3+{2\over 3}s_4\nn\\
s_5&=&-{1\over 3}s_3+{1\over 3}s_4\nn\\
s_6&=&-s_1+s_2. \eea
Therefore, the lagrangian at $l^4$ can be written as
\be \hat{L}=\sum_{i=1}^3 A_{i} t_i+\sum_{i=1}^4 B_{i} s_i +{\cal
O}(\Lambda^4), \label{lagrangian} \ee
where the constant coefficients $A,B$ are determined by the
requirement that the supersymmetric variation of $\hat L$ be a total
derivative, up to $\nab v$ terms.

At linear order in the fermions, the supersymmetric variation of
the correction to the Lagrangian, igoniring the $\nab v$ terms, consists of terms of the form
$\nabla \Lambda X^3$, $\Lambda\nabla X X^2$. Explicitly, one can
write down the following structures (suppressing spinor indices and the supersymmetry 
parameter $v$)

\noindent $\nabla \Lambda X^3$:
\bea w_1&:=&(\gamma_{ab}\otimes\sigma^{a'b'c'}\nabla_c\Lambda_d)
X^a{}_{ea'}X^{eb}{}_{b'}X^{cd}{}_{c'}\nn\\
w_2&:=&(\gamma_{ab}\otimes\sigma^{a'}\nabla_c\Lambda_d)
X^a{}_{ea'}X^{eb}{}_{b'}X^{cdb'}\nn\\
w_3&:=&(\gamma_{ab}\otimes\sigma^{a'}\nabla_c\Lambda_d)
X^{ac}{}_{a'}X^{eb}{}_{b'}X_e{}^{db'}\nn\\
w_4&:=&(\sigma^{a'}\nabla_a\Lambda_b)
X^a{}_{ea'}X^{eg}{}_{b'}X_g{}^{bb'}\nn\\
w_5&:=&(\sigma^{a'}\nabla_a\Lambda_b)
X^{abb'}X^{eg}{}_{a'}X_{egb'}\nn\\
w_6&:=&(\sigma^{a'}\nabla_a\Lambda_b)
X^{ab}{}_{a'}X^{eg}{}_{b'}X_{eg}{}^{b'}\nn\\
w_7&:=&(\gamma_{ab}\otimes\sigma^{a'}\nabla_c\Lambda_d)
X^{ac}{}_{b'}X^{eb}{}_{a'}X_e{}^{db'}\nn\\
w_8&:=&(\gamma_{ab}\otimes\sigma^{a'}\nabla_c\Lambda_d)
X^{ac}{}_{b'}X^{ed}{}_{a'}X_e{}^{bb'}\nn\\
w_9&:=&(\sigma^{a'}\nabla_a\Lambda_b)
X^a{}_{eb'}X^{eg}{}_{a'}X_g{}^{bb'}\nn\\
w_{10}&:=&(\gamma_{ab}\otimes\sigma^{a'b'c'}\nabla_c\Lambda_d)
X^{cb}{}_{a'}X^{a}{}_{eb'}X^{ed}{}_{c'} \eea

\noindent $\Lambda\nabla X X^2$:
\bea z_1&:=&(\sigma^{a'}\Lambda^a)
\nabla_a X_{bca'}X^{bdb'}X^{c}{}_{db'}\nn\\
z_2&:=&(\gamma^{ab}\otimes\sigma^{a'}\Lambda^c)
\nabla_d X_{bca'}X_{aeb'}X^{deb'}\nn\\
z_3&:=&(\sigma^{a'}\Lambda_a)
\nabla_b X_{cda'}X^{abb'}X^{cd}{}_{b'}\nn\\
z_4&:=&(\gamma^{ab}\otimes\sigma^{a'}\Lambda^c)
\nabla_b X_{dea'}X_c{}^{db'}X^{e}{}_{ab'}\nn\\
z_5&:=&(\sigma^{a'}\Lambda^a)
\nabla_a X_{bc}{}^{b'}X^{bd}{}_{a'}X^{c}{}_{db'}\nn\\
z_6&:=&(\gamma^{ab}\otimes\sigma^{a'}\Lambda^c)
\nabla_d X_{bcb'}X_{aea'}X^{deb'}\nn\\
z_7&:=&(\gamma^{ab}\otimes\sigma^{a'}\Lambda^c)
\nabla_b X_{deb'}X_{aca'}X^{deb'}\nn\\
z_8&:=&(\sigma^{a'}\Lambda_a)
\nabla_b X_{cdb'}X^{aba'}X^{cdb'}\nn\\
z_9&:=&(\gamma^{ab}\otimes\sigma^{a'}\Lambda^c)
\nabla_d X_{bcb'}X_{a}{}^{eb'}X^{d}{}_{ea'}\nn\\
z_{10}&:=&(\gamma^{ab}\otimes\sigma^{a'}\Lambda^c)
\nabla_c X_{de}{}^{b'}X^{d}{}_{aa'}X^{e}{}_{bb'}\nn\\
z_{11}&:=&(\gamma^{ab}\otimes\sigma^{a'}\Lambda^c)
\nabla_b X_{de}{}^{b'}X_{acb'}X^{de}{}_{a'}\nn\\
z_{12}&:=&(\gamma^{ab}\otimes\sigma^{a'b'c'}\Lambda^c)
\nabla_d X_{bca'}X_{aeb'}X^{de}{}_{c'}\nn\\
z_{13}&:=&(\gamma^{ab}\otimes\sigma^{a'b'c'}\Lambda^c)
\nabla_c X_{dea'}X^{d}{}_{ab'}X^{e}{}_{bc'}\nn\\
z_{14}&:=&(\gamma^{ab}\otimes\sigma^{a'b'c'}\Lambda^c)
\nabla_b X_{dea'}X_{acb'}X^{de}{}_{c'}\nn\\
z_{15}&:=&(\gamma^{ab}\otimes\sigma^{a'}\Lambda^c)
\nabla_b X_{deb'}X_{aea'}X_{c}{}^{db'}\nn\\
z_{16}&:=&(\gamma^{ab}\otimes\sigma^{a'}\Lambda^c)
\nabla_b X^{deb'}X_{aeb'}X_{cda'}\nn\\
z_{17}&:=&(\gamma^{ab}\otimes\sigma^{a'b'c'}\Lambda^c)
\nabla_b X^{de}{}_{a'}X_{aeb'}X_{cdc'}\nn\\
z_{18}&:=&(\sigma^{a'}\Lambda_a)
\nabla_b X_{cdb'}X^{abb'}X^{cda'}\nn\\
z_{19}&:=&(\gamma^{ab}\otimes\sigma^{a'}\Lambda^c)
\nabla_b X_{dea'}X^{deb'}X_{acb'}\nn\\
z_{20}&:=&(\sigma^{a'b'c'}\Lambda_a) \nabla_b
X_{cda'}X^{ab}{}_{b'}X^{cd}{}_{c'}. \eea
As before, not all of these structures are linearly independent.
Only six of the $w$ and fourteen of the $z$-terms are, as one can
see by counting the number of spinors $[(1)\times(0001)]$ in the
decomposition of the tensor products
\be [(5)\times(0010)]\otimes[(4)\times(1000)]^{3\otimes_s} \ee
and
\be [(3)\times(0010)]\otimes[(6)\times(1000)]
\otimes[(4)\times(1000)]^{2\otimes_s}. \ee
Indeed, taking into account (\ref{rc}) and the identity
\be \gamma_{[ab}\Lambda_{c]}=0, \ee
one can derive the following relations
\bea
w_7&=&-w_2+w_3-w_4+{1\over 2}w_6\nn\\
w_8&=&w_2+w_7\nn\\
w_9&=&-2w_4+w_5+{1\over 2}w_6\nn\\
w_{10}&=&{1\over 2}w_1 \label{plirpa} \eea
and
\bea
z_{15}+z_{16}&=&2z_5-z_6+z_7-z_9+z_{11}\nn\\
z_{15}-z_{16}&=&z_{10}\nn\\
z_{17}&=&{1\over 2}z_{13}\nn\\
z_{19}&=&-z_1+z_2+z_4\nn\\
z_{18}&=&-z_5+z_8+z_9+z_{10}-z_{11}+z_{15}\nn\\
2z_{20}&=&z_{12}-z_{13}-z_{14}-z_{17}. \label{plirpb} \eea
The last two lines in equation (\ref{plirpb}) above are derived by
noting that
\be (\gamma^{[ea}\sigma_{a'}\Lambda_e) \nabla_b
X_{c}{}^{d|b'}X_{a}{}^{ba'}X^{|c]}{}_{db'}=0 \ee
and
\be (\gamma^{[ea|}\sigma^{a'b'c'}\Lambda_e) \nabla_b
X_{c}{}^{|d|}{}_{a'}X_{a}{}^{b}{}_{b'}X^{|c]}{}_{dc'}=0. \ee
In addition one has to allow for integrations by parts. Up to
total derivatives, one has the following relations
\bea
w_1&=&2z_{17}\nn\\
w_2&=&z_4-z_{15}\nn\\
w_3&=&z_2-z_7\nn\\
w_4&=&-z_1-z_8\nn\\
w_5&=&-z_3-z_{18}\nn\\
w_6&=&-2z_8\nn\\
w_7&=&z_6-z_{19}\nn\\
w_8&=&z_9-z_{11}\nn\\
w_9&=&-z_3-z_5\nn\\
w_{10}&=&-z_{12}-z_{14} \label{plirpc} \eea
and
\bea
z_1&=&-2z_5\nn\\
z_2&=&z_6+z_{10}\nn\\
z_2&=&w_4-z_{16}\nn\\
z_4&=&w_9-z_9\nn\\
z_4&=&-w_7+z_{10}\nn\\
z_5&=&-w_4-z_{18}\nn\\
z_6&=&w_4-z_{15}\nn\\
z_7&=&{1\over 2}w_6\nn\\
z_{10}&=&-w_3-z_{15}\nn\\
z_{11}&=&w_5-z_{19}\nn\\
z_{12}&=&z_{17}\nn\\
z_{13}&=&2z_{12}\nn\\
z_{13}&=&w_{10}+z_{17}\nn\\
z_{16}&=&-w_8\nn\\
z_{20}&=&0. \label{plirpd} \eea
Equations (\ref{plirpa}, \ref{plirpb}, \ref{plirpc}, \ref{plirpd})
above define an overdetermined system of thiry-five equations for
thirty unknowns. It turns out however, that
 only twenty-four of the equations are linearly independent.
Therefore all $w$ and $z$-terms are expressible in terms of six
independent structures, which we can take to be $z_{1-4},
z_7,z_{12}$. Explicitly,
\bea
w_1&=&2z_{12}\nn\\
w_2&=&{1\over 2}z_1+z_2+z_4-z_7\nn\\
w_3&=&z_2-z_7\nn\\
w_4&=&-z_1+z_7\nn\\
w_5&=&-{3\over 2}z_1-z_3+z_7\nn\\
w_6&=&2z_7\nn\\
w_7&=&{1\over 2}z_1-z_4\nn\\
w_8&=&z_1+z_2-z_{7}\nn\\
w_9&=&{1\over 2}z_1-z_3\nn\\
w_{10}&=&z_{12} \label{fa} \eea
and
\bea
z_5&=&-{1\over 2}z_1\nn\\
z_6&=&-{1\over 2}z_1+z_2\nn\\
z_8&=&-z_7\nn\\
z_9&=&{1\over 2}z_1-z_3-z_{4}\nn\\
z_{10}&=&{1\over 2}z_1\nn\\
z_{11}&=&-{1\over 2}z_1-z_2-z_3-z_4+z_7\nn\\
z_{13}&=&2z_{12}\nn\\
z_{14}&=&-2z_{12}\nn\\
z_{15}&=&-{1\over 2}z_1-z_2+z_7\nn\\
z_{16}&=&-z_1-z_2+z_7\nn\\
z_{17}&=&z_{12}\nn\\
z_{18}&=&{3\over 2}z_1-z_7\nn\\
z_{19}&=&-z_1+z_2+z_4\nn\\
z_{20}&=&0. \label{fb} \eea
Again we note that the group-theoretic relations (\ref{plirpa},
\ref{plirpb}) are all implied by the partial integration equations
(\ref{plirpc}, \ref{plirpd}).

Next we compute the supersymmetric variation of the lagrangian
(\ref{lagrangian}). One finds, up to total derivatives and terms containing $\nab v$,
\bea
\delta t_1&=&-4i w_5\nn\\
\delta t_2&=&-4i w_6\nn\\
\delta t_3&=&-4i w_4\nn\\
\delta s_1&=&{1\over 2}(w_3+w_4-z_1+z_2)\nn\\
\delta s_2&=&{1\over 2}(w_2+w_5-z_3+z_{19})\nn\\
\delta s_3&=&{1\over 2}(w_4-w_7+w_8-w_9-w_{10}+z_6-z_9-z_{12})\nn\\
\delta s_4&=&{1\over
2}(w_1+w_2-w_5+w_6+z_7-z_8-z_{11}-z_{14}+z_{18}). \eea
Using (\ref{fa},\ref{fb}) it is straightforward to see that the
vanishing of the variation of the lagrangian implies the following
relations for the coefficients in (\ref{lagrangian}),
\bea
A_1&=&{3i\over 4}B_4\nn\\
A_2&=&0\nn\\
A_3&=&-{3i\over 4}B_4\nn\\
B_1&=&0\nn\\
B_2&=&-3B_4\nn\\
B_3&=&2B_4. \eea
These results lead to the Lagrangian \eq{alle}.
%

%
%%%%%%%%%%%%%%%%%%%%%%%%%%%%%%%%%%%%%%%%%%%%
%
% Bibliography

%
%%%%%%%%%%%%%%%%%%%%%%%%%%%%%%%%%%%%%%%%%%%%

\end{document}